%% file: paper_O4a.tex
\documentclass[%
aps,
prd,
amssymb,
amsmath,
amsfonts,
eqsecnum,
showpacs,
nofootinbib,
floatfix,
twocolumn,
twoside,
a4paper,
superscriptaddress,
longbibliography
]{revtex4-2}

\input{UsePackages}
\usepackage{graphicx}
\usepackage{dcolumn}
\usepackage{bm}
\usepackage{tabularx}
\usepackage{multirow}
\usepackage{capt-of}
\usepackage{color}
\usepackage{url}
\usepackage[utf8]{inputenc}
\usepackage{placeins}

\usepackage{ulem}

\usepackage{graphicx}
\usepackage{booktabs}
\usepackage{adjustbox}
\usepackage{amsmath}

\usepackage{natbib}

\usepackage[nolist,nohyperlinks]{acronym}
\usepackage{xspace}
\usepackage[colorlinks]{hyperref}
\usepackage{orcidlink}

\usepackage[caption=false]{subfig}

\usepackage{float}

\DeclareMathAlphabet{\mathbfsf}{\encodingdefault}{\sfdefault}{bx}{sl}

\newcommand{\be}{\begin{equation}}
\newcommand{\ee}{\end{equation}}
\newcommand{\bea}{\begin{eqnarray}}
\newcommand{\eea}{\end{eqnarray}}

\newcommand{\phT}{\textsc{IMRPhenomT}\xspace}

\newcommand{\phTHM}{\textsc{IMRPhenomTHM}\xspace}
\newcommand{\phTPHM}{\textsc{IMRPhenomTPHM}\xspace}

\definecolor{dodgerblue}{HTML}{1E90FF}
\definecolor{viennared}{HTML}{DA0A14}
\definecolor{ctorange}{HTML}{FF6C0C}
\definecolor{granadagreen}{HTML}{078931}
\definecolor{wales}{HTML}{ff0038}
\definecolor{valenciacfred}{HTML}{ee3524}
\definecolor{barcelonafcgold}{HTML}{edbb00}
\definecolor{jam}{HTML}{A50B5E}
\definecolor{austriawien}{HTML}{441678}

\AtBeginDocument{%
  \hypersetup{
    citecolor=dodgerblue,
    linkcolor=dodgerblue,   
    urlcolor=dodgerblue}}

\newcommand{\soft}[1]{\textsc{#1}}

\newcommand{\PYTHON}{\soft{Python}\xspace}

\newcommand{\GWPY}{\soft{GWpy}\xspace}

\newcommand{\PEAUTOMATOR}{\soft{PEAutomator}\xspace}
\newcommand{\GWTCFOUR}{GWTC-4.0\xspace}
\newcommand{\GWTCFIVE}{GWTC-5.0\xspace}

\newcommand{\IMRPhenomXHM}{\soft{IMRPhenomXHM}\xspace}

\newcommand{\IMRPhenomXPNR}{\soft{IMRPhenomXPNR}\xspace}
\newcommand{\IMRPhenomT}{\soft{IMRPhenomT}\xspace}
\newcommand{\IMRPhenomTHM}{\soft{IMRPhenomTHM}\xspace}
\newcommand{\IMRPhenomTPHM}{\soft{IMRPhenomTPHM}\xspace}

\newcommand{\IMRPhenomTHMtwozero}{\soft{IMRPhenomTHM\_20}\xspace}
\newcommand{\IMRPhenomTPHMtwozero}{\soft{IMRPhenomTPHM\_20}\xspace}

\newcommand{\SEOBNRFIVEPHM}{\soft{SEOBNRv5PHM}\xspace}

\def\nr#1{numerical relativity
 (NR)#1\gdef\nr{NR}}
\def\bh#1{black-hole
 (BH)#1\gdef\bh{BH}}
\def\bbh#1{binary black hole#1
 (BBH#1)\gdef\bbh{BBH}}
\def\pn#1{post-Newtonian (PN)#1\gdef\pn{PN}}
\def\imr#1{inspiral-merger-ringdown (IMR)#1\gdef\imr{IMR}}
\def\eob#1{effective-one-body
 (EOB)#1\gdef\eob{EOB}}
\def\td#1{time domain (TD)#1\gdef\td{TD}}
\def\fd#1{frequency-domain (FD)#1\gdef\fd{FD}}

\newcommand{\UIB}{Departament de F\'isica, Universitat de les Illes Balears, IAC3 -- IEEC, Crta. Valldemossa km 7.5, E-07122 Palma, Spain}

\newcommand{\ICE}
{Institut de Ci\`encies de l'Espai (ICE, CSIC), Campus UAB, Carrer de Can Magrans s/n, 08193 Cerdanyola del Vall\`es, Spain}

\allowdisplaybreaks[1]

\begin{document}

\title[ML]
{A stepping stone toward detecting gravitational wave memory: a cumulative analysis with the full $(\ell=2,m=0)$ spherical harmonic using events from \GWTCFOUR and \GWTCFIVE}

\author{Maria Rossell\'o-Sastre\,\orcidlink{0000-0002-3341-3480}}
\affiliation{\UIB}

\author{Sascha Husa\,\orcidlink{0000-0002-0445-1971}}
\affiliation{\ICE}
\affiliation{\UIB}

\author{Yumeng Xu\,\orcidlink{0000-0001-8697-3505}}
\affiliation{\UIB}

\author{Jorge Valencia\,\orcidlink{0000-0003-2648-9759}}
\affiliation{\UIB}

\author{Joan Llobera-Querol\,\orcidlink{0000-0003-3322-6850}}
\affiliation{\UIB}

\author{Antoni Ramos-Buades\,\orcidlink{0000-0002-6874-7421}}
\affiliation{\UIB}

\date{\today}

\begin{abstract}
We perform Bayesian model selection to test for the presence of the $(\ell=2,m=0)$ spherical harmonic mode in gravitational wave events that have previously been identified as binary black hole mergers. As our signal model we use the quasi-circular, non-precessing \IMRPhenomTHMtwozero waveform model, which includes the oscillatory and displacement memory contributions. Including the oscillatory component of the $(2,0)$ mode increases the signal-to-noise ratio and evidence for this mode, compared to testing only for the presence of gravitational wave memory. Our analysis thus constitutes a natural stepping stone toward detecting gravitational wave memory. We perform our analysis for the binary black hole signals
identified in the \GWTCFOUR catalog, and for selected \GWTCFIVE events. In our Bayesian model comparison we find a cumulative $\log_{10}\mathcal{B}=1.38\pm0.79$ in favor of the presence of the $(2,0)$ mode for the  \GWTCFOUR catalog. We also stack the signal-to-noise ratio 
of the full $(2,0)$ mode and of its individual contributions, obtaining results consistent with previous studies and reaching $\mathrm{SNR}_{\mathrm{memory}} = 0.89^{+0.29}_{-0.11}$ after approximately $7.5$ months of O4a observations. In addition, we study the precessing candidate GW241127\_061008, and find no additional evidence for the $(2,0)$ mode when precession is included in \IMRPhenomTPHMtwozero. Overall, our results provide an assessment of the observational support for the $(2,0)$ mode in current gravitational wave data and allow us to discuss prospects for its future detection. We find that decisive statistical evidence will likely require a larger catalog, with an optimistic estimated number of events of $N_{\mathrm{events}} = 166^{+82}_{-55}$, based on the specific assumptions adopted in this work. We also expect that decisive evidence will require a more extensive waveform systematics study.
\end{abstract}


\maketitle

\section{Introduction}
\label{sec:Introduction}

Accurate inference of source parameters in gravitational wave (GW) data analysis
require waveform models that capture as many measurable signal contributions as possible in the detector network. The emission of GWs  is commonly described in terms of a decomposition into a sum of spin-weighted spherical harmonic modes $(\ell,m)$, which provide a natural basis for describing the angular structure of the radiation \cite{10.1063/1.1705135}.
Here we are interested in compact binary coalescence, in particular of orbiting binary black hole (BBH) systems. In this case, the dominant harmonic is the $(\ell=2,m=\pm2)$ mode when the component spins are aligned with the orbital angular momentum. For misaligned binaries this statement more naturally holds in a co-precessing frame \cite{PhysRevD.84.024046}.
%
However, subdominant modes can play a significant role for asymmetric or highly inclined systems \cite{Mills:2020thr}, where they help break degeneracies among intrinsic and extrinsic parameters and thereby enhance measurement accuracy \cite{CalderonBustillo:2015lrt, Shaik:2019dym, Graff:2015bba,CalderonBustillo:2020kcg,CalderonBustillo:2022ldv}. 
Waveform models incorporating higher-order modes have been used systematically in the analysis of GW data since the third observing run (O3). To date, LIGO–Virgo–KAGRA (LVK) \cite{Aasi_2015,Acernese_2015,kagra_2021} 
observations have provided evidence for higher-order modes in the asymmetric events GW190412 and GW190814 \cite{LIGOScientific:2020stg,Hoy:2021dqg,LIGOScientific:2020zkf}. These measurements are dominated by the $(\ell=3,m=\pm3)$ multipole, as analyses of both events found this to be the only measurable subdominant mode. In the GWTC-2.1 sample, no event exhibited measurable higher-order-mode content beyond $\ell=3$, although a few additional events showed only marginal indications of the $(3,\pm3)$ mode.

Here we are interested in the $(\ell=2,m=0)$ mode, which contains the main contribution from the non-oscillatory displacement memory \cite{Favata:2009ii,Favata:2010zu,Mitman:2020pbt}, but also an oscillatory ringdown component. 
The memory arises from the cumulative emission of gravitational radiation throughout the evolution of the binary and manifests as a secular, permanent change between the initial and final asymptotic states of the spacetime. Unlike the oscillatory modes ($m \neq 0$), which encode the transient quasi-normal relaxation of the remnant black hole, the displacement memory instead is sourced by the
net flux of energy and momentum carried away by the GW emission. As such, it provides complementary information about the non-linear dynamics of the coalescence and offers a direct signature of the strong-field, radiative regime of general relativity (GR) \cite{Polnarev, Blanchet:1992br, PhysRevLett.67.1486}. 
Moreover, the distinct dependence of the $(2,0)$ mode on the inclination angle, opposite to that of the dominant $(2,\pm2)$ modes, helps break the distance–inclination degeneracy \cite{Xu:2024ybt, Gasparotto:2023fcg,CalderonBustillo:2020fyi}. 
Including the oscillatory component of the $(2,0)$ mode increases the signal-to-noise ratio (SNR) for this mode (compare e.g.\cite{Rossello-Sastre:2024zlr}), which will facilitate its detection and increases its relevance for parameter estimation (PE) of very high SNR events. Investigating the evidence for this mode thus constitutes a natural stepping stone for the detection of gravitational wave memory.
 

Motivated by these considerations and by the high number of GW events already observed by the LVK 
we perform a systematic reanalysis of BBH events detected
during the first two parts of the fourth observing run (O4a and O4b), and published as the 
\GWTCFOUR \cite{LIGOScientific:2025hdt,LIGOScientific:2025yae,LIGOScientific:2025slb,ligo_scientific_collaboration_and_virgo_2025_17014085,LIGO:2024kkz} and \GWTCFIVE \cite{LIGOScientific:2026sit,LIGOScientific:2026ifv,LIGOScientific:2026wfs,LIGOScientific:2026uyd,LIGOScientific:2026ctl} editions of the LVK GW 
transient catalog.
Building on the PE 
framework documented in our first analysis of the \GWTCFOUR catalog \cite{Xu:2025ajj}, we use the phenomenological waveform models 
to analyze the set of 84 BBH detections reported in \GWTCFOUR and a selection of 6 high significance BBH events reported in \GWTCFIVE. They are chosen based on high network SNR values, or high effective spin parameter, $|\chi_{\text{eff}}|$ \cite{LIGOScientific:2025rid, LIGOScientific:2025brd,LIGOScientific:2026wfs}, which enhances the $(2,0)$ mode, especially the amplitude of the memory contribution in the case of high positive spin components. 

For our analysis we are using the \IMRPhenomT family of time-domain phenomenological waveform models, restricted to the quasi-circular sector  \cite{Estelles:2020osj,Estelles:2020twz,Estelles:2021gvs}. The \IMRPhenomT family has been extended to  
include the $(\ell=2,m=0)$ spherical harmonic mode with the oscillatory ringdown and the displacement memory contributions
\IMRPhenomTHMtwozero \cite{Rossello-Sastre:2024zlr} for non-precessing, quasi-circular BBHs; and further to spin precession \IMRPhenomTPHMtwozero \cite{Rossello-Sastre:2025dep}. 

We quantify the evidence for the presence of the $(2,0)$ mode in individual events via Bayesian model comparison, assessing whether including the mode yields a statistically significant improved description of the observed data, 
and evaluate its cumulative detectability across the population by stacking event level statistics and computing a population-level SNR. In contrast to approaches that isolate only the displacement memory contribution, we perform a per event comparison between complete inspiral–merger–ringdown phenomenological models with and without the full $(2,0)$ mode, using O4 data and incorporating measured power spectral densities (PSDs) and calibration uncertainties. The resulting Bayes factors are then combined across events. 

The paper is organized as follows. We start with a review of detectability prospects for GW memory in Sec.~\ref{sec:memory_detectability}. In Sec.~\ref{sec:models}, we briefly describe the waveform model used in this work. In Sec.~\ref{sec:methods}, we outline the treatment of the data, and the methods used in the analysis. In Sec.~\ref{sec:results}, we present the results of the reanalysis of the dataset using the \IMRPhenomTHMtwozero model to include the $(\ell=2,m=0)$ mode. Finally, we summarize our findings and conclude in Sec.~\ref{sec:conclusions}.

Throughout this paper, component masses are denoted by $m_i$. We define the mass ratios $q = m_2/m_1 \leq 1$ and $Q=1/q$, and the total mass $M = m_1 + m_2$. Masses are reported in the detector frame. The components of the dimensionless spin vectors in the direction of the orbital momentum 
(our $z$-axis orthogonal to the orbital plane) are denoted by $\chi_i = S_i^z/m_i^2$.

\section{Prospects for GW memory detectability}
\label{sec:memory_detectability}
The detectability of GW memory has been the subject of extensive investigation, spanning analytical estimates, population-level forecasts, and Bayesian inference studies. Early observational analyses searched for memory signatures in existing GW catalogs, including GWTC-1 \cite{Khera:2020mcz, Hubner:2019sly}, GWTC-2 \cite{Zhao:2021hmx, Hubner:2021amk}, and GWTC-3 \cite{Cheung:2024zow}, but did not yield conclusive detections. Foundational work established that detection with second-generation detectors is most naturally achieved through stacking multiple events. In particular, \cite{Lasky:2016knh} showed that hundreds of GW150914-like signals at Advanced LIGO design sensitivity would be required for a confident detection using frequentist SNR 
Subsequent population-based forecasts, incorporating updated BBH distributions and detector sensitivities, predicted cumulative SNRs of order $\sim 3$ after several years of observation, reinforcing the conclusion that detection is more likely through population stacking than via individual events \cite{Boersma:2020gxx}. More studies, including the effects of spin memory and next generation detector networks, have further refined this picture: while a second generation network operating at O4/O5 sensitivities may achieve population level detection of displacement memory, third generation observatories such as Cosmic Explorer \cite{Reitze:2019iox} are expected to enable detections in individual events. 
In contrast, the detection of spin memory is likely to still rely on stacking even in these advanced scenarios \cite{Grant:2022bla}. 

In a recent work \cite{Mitman:2026zfg}, the authors use hierarchical Bayesian inference on the \GWTCFIVE  
catalog of BBH observations to constrain the memory enhancement factor across the population, finding a value consistent with the GR prediction 
and forecasting that roughly 2000 detections would be needed to measure a non-zero memory signal at the $1\sigma$ level. Complementary analyses have explored the detectability of memory at the level of individual events with current detectors. In particular, \cite{Chen:2024ieh} performed Bayesian model selection studies for signals observed with O4 sensitivity and third generation detectors. For an optimally oriented (edge-on) injection with a network SNR of 44.7 at O4 sensitivity, repeated injection–recovery analyses across multiple noise realizations yielded Bayes factors clustered around zero, remaining below detection thresholds and providing no statistically significant preference for either the memory or no memory hypothesis. 

In the space-based context, early theoretical studies also highlighted the strong potential of LISA \cite{LISA:2024hlh} 
for per event memory detection in massive black hole binaries (MBHBs) \cite{Favata:2009ii,Favata:2010zu}. This expectation has been confirmed and substantially refined by recent work. Using time domain modeling of the LISA response, \cite{Inchauspe:2024ibs} demonstrated that the memory signal induces a resolvable, non-oscillatory component in the strain, with SNRs sufficient for detection in a subset of MBHB mergers over the mission lifetime. Building on this, \cite{Cogez:2026frh} carried out a full Bayesian parameter estimation analysis, showing that the memory amplitude can be constrained with meaningful precision and that waveform models including memory are robustly supported by the data. Finally, \cite{Zosso:2026czc} developed a rigorous framework for memory detection, introducing parametrized descriptions of the memory growth and establishing hypothesis-testing strategies based on model selection, thereby defining the conditions for confident detection. In contrast, searches for bursts with memory in pulsar timing array data have thus far yielded only upper limits. For example, analyses of the NANOGrav 12.5-year dataset report no statistically significant evidence for memory signals, with Bayes factors of $\sim 2.8$ for models including memory in addition to a stochastic GW background \cite{NANOGrav:2023vfo, Tomson:2025ixn}. 


\section{Waveform model}\label{sec:models}

The \phT family of time domain phenomenological models \cite{Estelles:2020twz,Estelles:2020osj,Estelles:2021gvs,Planas:2025feq,Rossello-Sastre:2024zlr,Rossello-Sastre:2025dep} offers an alternative to the computationally efficient frequency domain models. The flexibility of modeling effects in the time domain facilitates the inclusion of complicated new physical effects such as spin-precession \cite{Estelles:2021gvs}, orbital eccentricity \cite{Planas:2025feq}, or GW memory \cite{Rossello-Sastre:2024zlr}.  For this analysis, we employ the 
quasi-circular, non-precessing-spin \phTHM model  \cite{Estelles:2020osj,Estelles:2020twz}, which has been calibrated to numerical relativity (NR) 
and test-particle waveforms and it includes $(\ell,m)=\{(2,\pm2),(2,\pm1),(3,\pm3),(4,\pm4),(5,\pm5)\}$ multipoles; and the quasi-circular precessing-spin \phTPHM model \cite{Estelles:2021gvs}, which is based on the ``twisting-up'' approximation \cite{Schmidt:2014iyl} and has the same mode content in the co-precessing frame as the \phTHM model. 

To incorporate the memory effect in this analysis, we employ the phenomenological waveform model of the $(\ell=2, m=0)$ spherical harmonic mode, which captures both the dominant displacement memory contribution and the quasinormal mode (QNM) 
oscillations of the ringdown in this mode \cite{Rossello-Sastre:2024zlr}. This model is constructed in the time domain by fitting phenomenological \textit{Ansätze} to NR 
simulations and memory waveforms computed using the Bondi-Metzner-Sachs balance laws \cite{Ashtekar:2019viz,Compere:2019gft}. 
It is valid for BBH systems in quasi-circular 
orbits, covering mass ratios up to $Q=10$ and spin components aligned with the orbital angular momentum, $\chi_1,\chi_2\in[-1,1]$. In order to extend our description of the $(2,0)$ mode to include spin-precession, we apply the ``twisting-up'' approximation to the aligned-spin model of the $(2,0)$ mode and we include the displacement memory contribution in all the $\ell=2$ modes \cite{Rossello-Sastre:2025dep}. Both the aligned-spin and the precessing models are implemented within \phTHM and \IMRPhenomTPHM, respectively, in the \PYTHON package {\tt phenomxpy} \cite{phenomxpy}.

\section{Methods}\label{sec:methods}

The strain data is taken from the Gravitational Wave Open Science Center (GWOSC) \cite{LIGOScientific:2025snk} using the \GWPY package \cite{gwpy} with sample rate 4096 Hz. The data conditioning procedures, including the duration of the analyzed segments, the treatment of instrumental glitches, and the choice of the starting frequency used in the likelihood evaluation, follow the methodology described in the original analysis presented in \cite{Xu:2025ajj}. For completeness, we refer the reader in particular to Sec.~III B of this reference, as the inference framework and analysis configuration adopted in this work are identical to those employed in that study. The prior distributions used in the analysis are chosen to reflect standard astrophysical assumptions. Uniform priors are imposed on the redshifted component masses, the merger time, and the coalescence phase. The orientation of the binary and its sky location are assigned isotropic priors. For the luminosity distance, we adopt a prior corresponding to a constant merger rate density in comoving volume and source-frame time. For the spin components, we use \texttt{bilby}'s \cite{bilby} \texttt{AlignedSpin} with spin magnitudes uniformly distributed between 0 and 0.99 and spins constrained to be aligned with the orbital angular momentum.

Note that time domain waveforms which contain the displacement memory contribution are non-periodic signals. We need to take this into account when converting the waveforms to frequency domain. For this, we use the Fourier transform method {\tt gw-foutstep} \cite{Valencia:2024zhi}, which decomposes the waveform into a known sigmoid-like component (capturing the persistent offset and transformable analytically) plus a residual transient part that can be safely Fourier transformed numerically without artifacts.
This method avoids the spurious artifacts introduced by standard Fast Fourier Transform 
techniques applied to non-periodic signals. 

Because the present study requires the execution of multiple PE runs, minimizing manual interaction is essential in order to reduce the possibility of human error and ensure reproducibility. To this end, we employ the automated workflow tool \PEAUTOMATOR{} \cite{pe_automator}, which manages job submission, monitoring, and post-processing of the results.

\subsection{Setup and hypotheses}

Our goal is to assess the presence of the $(\ell=2,m=0)$ mode in the observed signals. This mode contains two physically distinct contributions: an oscillatory component associated with the QNM 
ringdown of the remnant black hole, and the displacement memory contribution. To quantify the evidence for this mode, we perform a paired Bayesian model comparison analysis for each event \cite{Thrane:2018qnx}. We consider two hypotheses:
\begin{itemize}
\item $\mathcal{H}_{20}$: \IMRPhenomTHMtwozero\ model with the $(2,0)$ mode (oscillatory ringdown + displacement memory),
\item
$\mathcal{H}_{\varnothing}$: \IMRPhenomTHM\ model without the $(2,0)$ mode.
\end{itemize}

For a given event $i$, we quantify the relative evidence for the presence of the mode through the Bayes factor 
\begin{equation}
\mathcal{B}_i \equiv \frac{p(d_i\!\mid\!\mathcal{H}_{20})}{p(d_i\!\mid\!\mathcal{H}_{\varnothing})},\qquad
\log_{10} \mathcal{B}_i=\log_{10} Z_i(\mathcal{H}_{20})-\log_{10} Z_i(\mathcal{H}_{\varnothing}),
\label{eq:Bi}
\end{equation}
where $Z_i(\mathcal{H})$ denotes the Bayesian evidence for the hypotheses $\mathcal{H}$ computed from the data $d_i$ of the corresponding event. The uncertainty on the log-10 Bayes factor is propagated from the uncertainties in the evidences via
\begin{equation}
\sigma\!\left[\log_{10} \mathcal{B}_i\right] 
= \sqrt{\sigma^2\!\left[\log_{10} Z_i(\mathcal{H}_{20})\right] 
+ \sigma^2\!\left[\log_{10} Z_i(\mathcal{H}_{\varnothing})\right]} .
\label{eq:BFerr}
\end{equation}
The evidences are computed using the \texttt{dynesty} nested-sampling algorithm as implemented in \texttt{bilby} \cite{bilby}. 
To assess the cumulative evidence for the $(2,0)$ mode across the population of detected events, we combine the individual Bayes factors assuming statistical independence between events. Under this assumption, the stacked Bayes factor is given by the sum of the  Bayes factor logarithms,
\begin{equation}
\log_{10} \mathcal{B}^{\mathrm{THM\_20/THM}}=\sum_{i=1}^{N}\log_{10} \mathcal{B}_i,
\label{eq:stackBF}
\end{equation}
where $N$ denotes the number of analyzed events. The corresponding uncertainty is obtained by adding the individual uncertainties in quadrature,
\begin{equation}
\sigma\!\left[\log_{10} \mathcal{B}^{\mathrm{THM\_20/THM}}\right] 
= \sqrt{\sum_{i=1}^N \sigma^2\!\left[\log_{10} \mathcal{B}_i\right]} .
\label{eq:stackBFerr}
\end{equation}

To interpret the Bayes factor results, we adopt the Jeffreys’ scale \cite{Jeffreys1998}, which provides a qualitative classification of the strength of evidence in favor of a given hypothesis. According to this scale, Bayes factors with $|\log_{10}\mathcal{B}| < 0.5$ are generally considered inconclusive, corresponding to evidence that is not worth more than a bare mention. Values in the range $0.5 \le |\log_{10}\mathcal{B}| < 1$ are interpreted as substantial evidence, while $1 \le |\log_{10}\mathcal{B}| < 2$ indicate strong evidence in favor of one model over the other. Finally, $|\log_{10}\mathcal{B}| \ge 2$ is typically regarded as decisive evidence. This classification applies symmetrically to negative values of $\log_{10}\mathcal{B}$, in which case the same levels of evidence favor the alternative hypothesis. Note that here we directly work with the Bayes factor instead of an odds ratio, since the parameter space that corresponds to the two models we are using for model selection is the same, and 
due to the smallness of the contribution of the $(2,0)$ mode to the signal no significant effect on detection is expected,
and we conclude from our results that there is also no significant effect on the estimated parameters.


\subsection{Stacking signal-to-noise ratio}
As an additional diagnostic, we compute the SNR associated with the $(2,0)$ mode and use it as a proxy for the strength of the evidence obtained from the Bayesian model comparison. This approach allows direct comparison with the methodology adopted in \cite{Grant:2022bla, Boersma:2020gxx}. In particular, we evaluate the SNR for three cases: the complete $(2,0)$ mode including both the oscillatory and displacement memory contributions; the displacement memory contribution alone; and the oscillatory component alone. For a set of independent events, the combined SNR is obtained by summing the individual SNRs in quadrature,

\begin{equation}
\label{eq:stackSNR}
    \rho^2=\sum_i\rho_i^2,
\end{equation}
where the SNR for a signal $h(t)$ is computed in frequency domain as
\begin{equation}
    \rho=\sqrt{\langle h|h\rangle}=\sqrt{4\int_0^{\infty}\frac{|\tilde{h}(f)|^2}{S_n(f)}df},
\end{equation}
with $S_n(f)$ the detector's PSD.

We restrict the $(2,0)$ mode SNR stacking analysis to the O4a event sample, excluding events selected from O4b in order to avoid introducing selection biases, since the O4b subset was not chosen in an unbiased or complete manner but rather according to specific selection criteria. Results are reported in Sec.~\ref{sec:stackingSNR}, where we find consistency with the predictions in \cite{Grant:2022bla, Boersma:2020gxx}.

\subsection{Bayes factor's scaling with the SNR}
\label{sec:BFscaling}
In order to compare the strength of the evidence across different PE analyses in a consistent way, it is useful to estimate how the Bayes factor changes as a function of the SNR. In the high-SNR approximation \cite{DelPozzo:2014cla}, and assuming that the two analyses are performed with the same priors and the same waveform mismatch, the Bayes factor is expected to scale as

\begin{equation}
\label{anBF}
\log_{10}\mathcal{B}\propto \frac{\rho^2}{2}\left(1-\mathrm{FF}^2\right),
\end{equation}
where $\rho$ is the SNR and FF is the fitting factor between the two waveforms, which quantifies the proximity of the target signal to a template manifold by measuring the reduction in recovered SNR when the signal is filtered with the best matching template from that manifold \cite{Owen:1995tm}. 
This allows us to rescale the Bayes factor measured for one run to estimate its expected value, and uncertainty, for another run with a different SNR. This scaling provides a simple way to extrapolate the significance of the results obtained for the louder events. Computing the ratio between two runs, assuming the same value for FF and that we use the same priors in both injections, we can compute the Bayes factor and its associated error as
\begin{equation}
\label{scaleBF}
    \log_{10}\mathcal{B}=\left(\frac{\rho}{\rho_0}\right)^2\log_{10}\mathcal{B}_0,\quad \sigma=\left(\frac{\rho}{\rho_0}\right)^2\sigma_0.
\end{equation}
With this expression  we can easily check the scaling of the Bayes factor with the SNR. This is 
presented in Sec.~\ref{sec:injections} using a set of PE, zero-noise model injections, for which we obtain results consistent with the expected scaling of these quantities.

\subsection{Sampler settings}
\label{sec:sampler_settings}
As previously mentioned, we employ the \texttt{dynesty} \cite{dynesty} sampler as implemented in \texttt{bilby} \cite{bilby}. We use the acceptance-walk method for the Markov Chain Monte Carlo 
evolution with {\tt naccept=60} and {\tt nlive=1000} with two independent seeds for each run. The stopping criterion of the sampler is controlled by the parameter $\Delta\ln Z$, which sets the threshold for the estimated remaining contribution of the unexplored prior volume to the total evidence. The sampler terminates once this residual contribution falls below the chosen tolerance. Therefore, reducing $\Delta\ln Z$ enforces a stricter convergence requirement, drives the nested sampling run further into the high likelihood region before stopping, and decreases the residual uncertainty in the final estimate of $\ln Z$. The default value in \texttt{bilby} is set to $\Delta\ln Z=0.10$, however here we adopt $\Delta\ln Z=0.05$ in order to obtain better-converged evidence estimates. Since the obtained Bayes factors are close to zero and, in most cases, compatible with zero within the statistical uncertainty, the evidence difference between the competing hypotheses is small and a more accurate determination of the residual error is desirable. A tighter stopping criterion yields smaller numerical uncertainties and more trustworthy Bayes factors at the expense of increased computational cost. However, the total wall time is not deterministic, since the early stages of the algorithm involve stochastic exploration of the parameter space and the number of iterations required to reach convergence can vary between runs, even for identical settings. Consequently, the observed runtime reflects both the systematic effect of the stopping criterion and the intrinsic variability of the sampling process, which can mask any strictly monotonic dependence on $\Delta\ln Z$. We provide in Appendix \ref{app:runtimes_dlogz} a more detailed discussion on the runtimes of all the events analyzed with both values of $\Delta\log Z$ and each model.



\subsection{Parameter estimation injection studies}
\label{sec:injections}

To validate our setup we perform a small set of injection studies.

\subsubsection{Numerical relativity injection}

First we perform a zero-noise NR injection using a quasi-circular, non-precessing, waveform from the Simulating eXtreme Spacetimes (SXS) public Catalog \cite{SpECWebsite,Boyle:2019kee,Scheel:2025jct}, SXS:BBH:2497 ($Q=2,\chi_{1}=\chi_{2}=0, N_{\mathrm{orbits}}=20$). We choose a non-equal mass simulation so that all the modes in the model have a non-zero contribution to the waveform. 
We recover the injected parameters with the model with and without the $(2,0)$ mode to be able to compute the Bayes factor between the two recoveries. We choose the same mode content in order to be able to see the biases that are introduced in the parameters only due to the omission of the $(2,0)$ mode. We select an inclination of $\pi/2$, as this is the optimal configuration for the $(2,0)$ mode and the network (H1 and L1) SNR of the injected signal is 37.6, 
with a $(2,0)$ mode SNR of 0.98. 
We pick the values in this range to keep the analysis realistic and comparable to the SNR values observed in the O4a and O4b observing runs. The injected parameters for this injection are listed in Tab.~\ref{table:pars_injections}. 
The posterior distributions are shown in Appendix \ref{app:nrinjection}. At this SNR and for this set of source parameters we observe no appreciable bias in the recovered intrinsic or extrinsic parameters when the $(2,0)$ mode is omitted from the recovery model. The Bayes factor between the two hypotheses in this case is $\log_{10}\mathcal{B}=0.48\pm0.12$, indicating only weak support for the model including the mode. These findings are consistent with our results for the O4a and O4b data: in both cases, omitting the $(2,0)$ mode does not lead to noticeable biases in the posterior distributions, and the Bayes factors remain of comparable magnitude.

\begin{table}[H]
\centering
\begin{tabular}{lcc}
\hline\hline
Parameter~~ & ~~NR injection~~ & ~~Model injection high (low) SNR \\
\hline
$q$                    & 2.00     & 1.03 \\
$M~[M_\odot]$          & 50       & 100 \\
$d_L~[\mathrm{Mpc}]$   & 400      & 140 (4000) \\
$\chi_1$               & 0.0      & 0.2 \\
$\chi_2$               & 0.0      & 0.2 \\
$\iota~[\mathrm{rad}]$ & $\pi/2$  & $\pi/2$ \\
$\phi~[\mathrm{rad}]$  & 1.20     & 5.43 \\
$\psi~[\mathrm{rad}]$  & 1.26     & 2.70 \\
$\alpha~[\mathrm{rad}]$& 2.60     & 2.77 \\
$\delta~[\mathrm{rad}]$& 0.53     & $-0.50$ \\
\hline\hline
\end{tabular}
\caption{Parameters of the NR injection (middle column) and model injections (right column). The only difference between the high and low SNR model injections is the luminosity distance, which is quoted in brackets for the low SNR injection. In all cases, we set the minimum frequency of the analysis to $f_{\mathrm{min}}=f_{\mathrm{ref}}=20$ Hz. 
}
\label{table:pars_injections}
\end{table}

\subsubsection{Model injections}
Furthermore, to assess the correct scaling of the Bayes factor with the SNR in the PE runs that have been performed in this work, we perform two zero-noise model injections: one at considerably high SNR and one at lower SNR, to check the scaling of the Bayes factor as a function of the SNR introduced in Sec. \ref{sec:BFscaling}. We choose model rather than NR injections for this purpose, in order to avoid waveform systematics between the injected waveform and the model used for recovery. First, we inject a signal at a high SNR of 435.3 and we compute the Bayes factor between the recovery with and without the $(2,0)$ mode in the model, obtaining the result: $\log_{10}\mathcal{B}=26.06\pm0.24$. Then, we perform the same injection at a much lower SNR of 15.3 and we compare the obtained Bayes factor, $\log_{10}\mathcal{B}=0.08\pm0.16$ with the analytical prediction based on Eq.~(\ref{anBF}). Using the expression in Eq.~(\ref{scaleBF}), with $\rho=15.3, \rho_0=435.3, \log_{10}\mathcal{B}_0=26.06, \sigma_0=0.24$, we obtain $\log_{10}\mathcal{B}\pm\sigma=0.03209\pm0.00030$. But taking into account that due to the sampler settings used, the run stopping criterion in this case is the default, $\Delta\ln Z=0.10$, the uncertainty will at least be $0.10$, leaving a final result of $\log_{10}\mathcal{B}\pm\sigma=0.03\pm0.10$, which is consistent with the result of the actual PE injection: $\log_{10}\mathcal{B}=0.08\pm0.16$. The parameters used for these injections are shown in Tab.~\ref{table:pars_injections} and the results are summarized in Tab.~\ref{table:BFsinjections}. 

In summary, at the low SNR considered here, there is no robust evidence for the presence of the $(2,0)$ mode. We emphasize that these zero-noise model injections only quantify modeling and sampling behavior. In real detector data, noise realizations add an additional stochastic contribution to the evidence. For Bayes factors of order unity or smaller such noise fluctuations can change the sign of $\log_{10}\mathcal{B}$ or produce values that are consistent with zero within the uncertainty, so small measured Bayes factors should be interpreted cautiously.

\begin{table}[htbp]
\centering
\begin{tabular}{ccc}
\hline\hline
                                       & SNR    & $\log_{10}\mathcal{B}\pm1\sigma$ \\
\hline
High SNR injection                     & 435.3 & $26.06\pm0.24$         \\
Low SNR injection                      & 15.3  & $0.08\pm0.16$          \\
Analytical value for low SNR injection & -      & $0.03\pm0.10$          \\
\hline\hline
\end{tabular}
\caption{Bayes factor scaling for two model injections at different SNR compared with the analytical estimate with Eq.~(\ref{anBF}).}
\label{table:BFsinjections}
\end{table}

\section{Results}
\label{sec:results}

We perform an analysis of the BBH events in \GWTCFOUR and the set of six events from \GWTCFIVE \cite{LIGOScientific:2026wfs} listed here (quoting the results obtained by the LVK analysis):

\begin{itemize}
    \item GW240920\_124024 is a high SNR$=37.0^{+0.1}_{-0.1}$ event, with some support for spin precession, $\chi_{\text{p}}=0.47^{+0.44}_{-0.30}$.
    \item GW240925\_005809 \cite{LIGOScientific:2026qqv} is a high SNR$=31.9^{+0.1}_{-0.1}$ event, the posterior however shows a bimodality in the inferred redshift. 
    \item GW241011\_233834 \cite{LIGOScientific:2025brd} is another high SNR$=35.8^{+0.1}_{-0.1}$ event. It shows
    unequal masses, high spins $\chi_{\text{eff}}=0.50^{+0.06}_{-0.04}$, and is the closest event among the candidates reported in \GWTCFIVE, $d_L=210^{+40}_{-40}$ Mpc.
    \item GW241110\_124123 \cite{LIGOScientific:2025brd} is a low SNR$=9.9^{+0.3}_{-0.5}$ event, with unequal masses and high spins $\chi_{\text{eff}}=-0.31^{+0.23}_{-0.18}$.
    \item GW241127\_061008 has high SNR$=31.2^{+0.1}_{-0.2}$, unequal masses, and support for spin precession $\chi_{\text{p}}=0.57^{+0.14}_{-0.21}$.
    \item GW250114\_082203 \cite{LIGOScientific:2025rid} has the highest SNR among observed GW events to date, SNR$=76.9^{+0.0}_{-0.1}$, providing a particularly interesting case for searching for subdominant features in the waveforms. 
\end{itemize}

We use the \IMRPhenomTHMtwozero and \IMRPhenomTHM waveform models in order to investigate the presence of the $(\ell=2,m=0)$ mode in the data. 
For the event GW241127\_061008, we also compare \IMRPhenomTPHMtwozero and \IMRPhenomTPHM, since its $\chi_{\mathrm{p}}$ value is significantly different from zero. We do not use precessing waveform models for the full event set because they are more computationally expensive than the aligned spin versions, and the majority of events show no decisive evidence for precession when analyzed with the waveform models included in \GWTCFOUR and \GWTCFIVE. 

\subsection{Bayes factors}
Using Eqs.~(\ref{eq:Bi}) and (\ref{eq:BFerr}), we compute the Bayes factors and their associated uncertainties for all tested hypotheses across the full set of considered events. Fig.~\ref{fig:logBF_perevent} shows the individual values of $\log_{10}\mathcal{B}^{\mathrm{THM\_20/THM}}$ for each BBH event in \GWTCFOUR and for the selected events from \GWTCFIVE (marked by the blue font color). The events are ordered by increasing SNR of the $(2,0)$ mode, which is encoded by the color of the marker edges, while the marker fill color represents the SNR of the full signal. Overall, the results indicate that for the majority of events the data do not provide strong evidence in favor of including the $(2,0)$ mode. Most measurements cluster around zero and their associated $1\sigma$ uncertainties typically overlap the null value, implying that the two waveform models are statistically indistinguishable. A mild tendency toward positive values is observed at increasing SNR, suggesting that the contribution of the $(2,0)$ mode may become marginally more relevant in louder signals, although the inferred Bayes factors remain small and correspond to weak evidence.

The labels highlighted in blue (red) correspond to events with $\log_{10}\mathcal{B}^{\mathrm{THM\_20/THM}} - 1\sigma > 10^{-3}$ ($\log_{10}\mathcal{B}^{\mathrm{THM\_20/THM}} + 1\sigma < -10^{-3}$), i.e., those showing the most robust, although still weak, preference for the inclusion (exclusion) of the $(2,0)$ mode; their values are reported in Tab.~\ref{table:log10BF20_events}. We adopt $10^{-3}$ rather than 0 to introduce a margin above the null evidence threshold $\mathcal{B}=1$, thereby avoiding the classification of marginal cases as positive due to numerical precision limits or fluctuations induced by the sampling procedure. 
The highest evidence for the presence of the $(2,0)$ mode is found for GW230814\_230901 \cite{LIGOScientific:2025cmm}, 
with substantial evidence according to Jeffreys' scale ($\log_{10}\mathcal{B}^{\mathrm{THM\_20/THM}}=0.63\pm0.09$). Consistently, the posterior distributions presented in Appendix~\ref{app:posteriors} show that, for all events, the inferred parameters are largely unaffected by the inclusion of the $(2,0)$ mode. The detectability of this mode is not solely driven by signal strength but also depends on additional source properties, such as the binary orientation and intrinsic parameters. To assess the agreement between the two waveform models and the data, in Appendix~\ref{app:posteriors} we also present the whitened strain for the two highest SNR events in the dataset: GW230814\_230901 and GW250114\_082203.

In our previous reanalysis of \GWTCFOUR \cite{Xu:2025ajj}, we identified systematic differences between \IMRPhenomTPHM and \IMRPhenomXPNR \cite{kxsf-23rr} for the event GW230814\_230901, which arise from their different mode content. In particular, these discrepancies can be traced to the absence of the $(3,\pm2)$ mode in the time domain phenomenological model. See Fig. 5 of Ref.~\cite{Xu:2025ajj} for the systematics in the mass ratio-inclination parameters across \IMRPhenomXPNR, \IMRPhenomTPHM and \SEOBNRFIVEPHM \cite{Pompili_2023,Ramos-Buades:2023ehm,Estelles:2025zah}, which also includes the $(3,\pm2)$ mode. This suggests that an important improvement to these waveform models would be the inclusion of the $(3,\pm2)$ multipole contribution. These differences in the posteriors can also be seen in the top left panel of Fig.~\ref{fig:posteriors_events} in Appendix~\ref{app:posteriors}, where the mixed posterior samples from \GWTCFOUR are plotted together with the results from the \IMRPhenomTHM and \IMRPhenomTHMtwozero runs. 
To assess whether \IMRPhenomTHMtwozero remains preferred over \IMRPhenomXHM (aligned-spin, frequency domain model which includes the $(3,\pm2)$ mode) under a consistent comparison, we perform a PE analysis using identical settings with the latter model. We obtain $\log_{10}\mathcal{B}^{\mathrm{THM\_20/XHM}} = 0.18 \pm 0.10$. While the evidence is 3.5 times lower than that obtained in the comparison against \IMRPhenomTHM, the result still provides mild support for the presence of the $(2,0)$ mode in the data.

\begin{widetext}
\begin{center}
\begin{figure}[H]
    \centering
    \includegraphics[width=\textwidth]{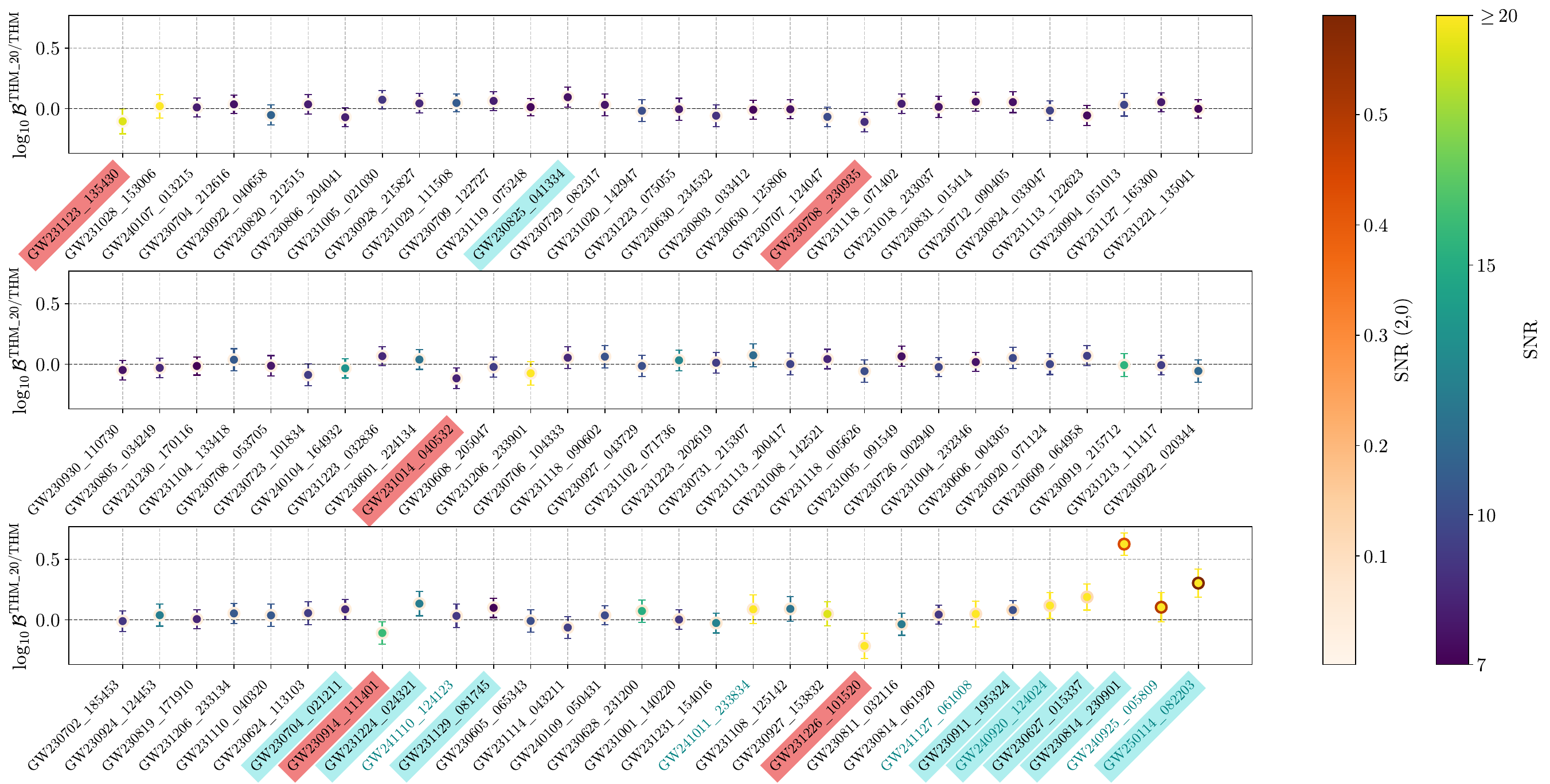}
    \caption{Log-10 Bayes factors for each individual event in increasing SNR of the $(2,0)$ mode order. Each point corresponds to the value of $\log_{10}\mathcal{B}^{\mathrm{THM\_20/THM}}$ with $1\sigma$ sampling uncertainty shown as vertical error bars. The color filling the markers represents the network SNR of the full signal for each event, as indicated by the rightmost colorbar; while the color of the border of the marker represents the network SNR of only the $(2,0)$ mode, as indicated in the leftmost colorbar. The horizontal dashed line marks the reference value $\log_{10} \mathcal{B}^{\mathrm{THM\_20/THM}}=0$, corresponding to no preference between the models. The events with $\log_{10}\mathcal{B}^{\mathrm{THM\_20/THM}}-1\sigma>10^{-3}$ are highlighted in blue and the ones with $\log_{10}\mathcal{B}^{\mathrm{THM\_20/THM}}+1\sigma<-10^{-3}$, in red. The events from \GWTCFIVE are marked with the color of the font label in blue.} 
    \label{fig:logBF_perevent}
\end{figure}
\end{center}
\end{widetext}

\begin{figure}[H]
    \centering
    \includegraphics[width=\columnwidth]{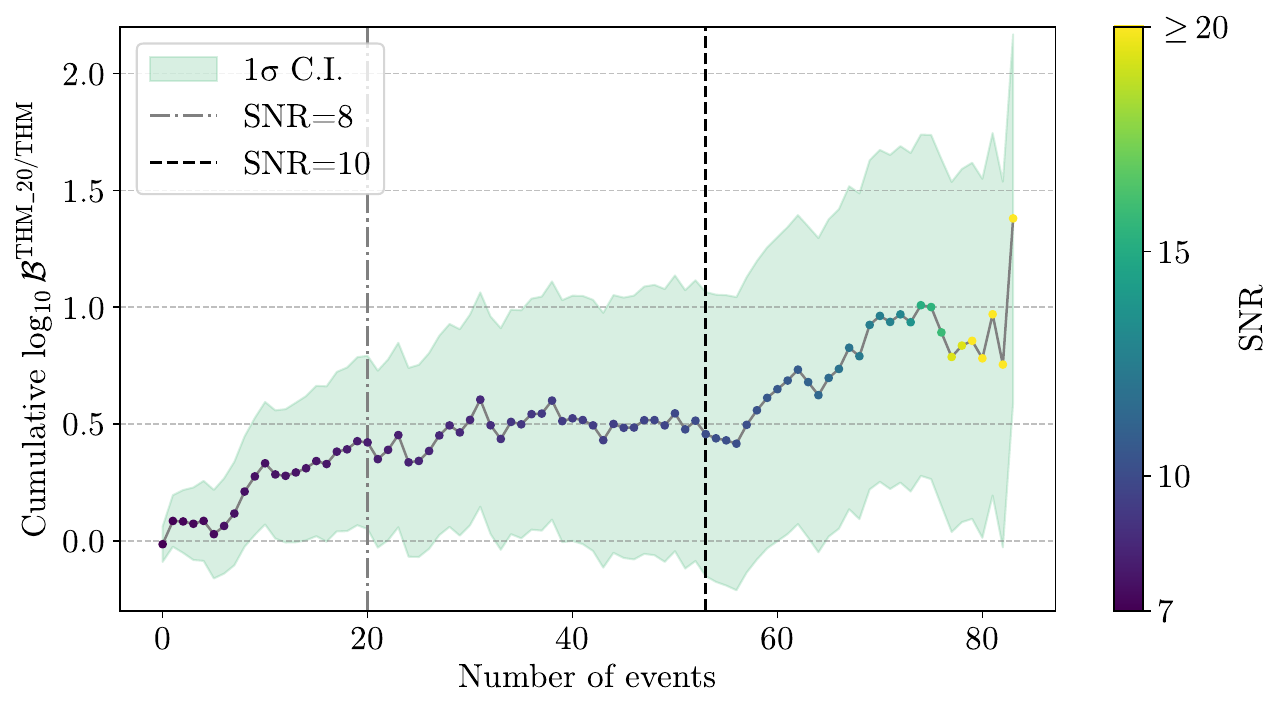}
    \caption{Cumulative sum of log-10 Bayes factors as a function of the number of events, in increasing SNR order for the BBH events in \GWTCFOUR. 
    The shaded region shows the propagated $1\sigma$ uncertainty and the color scale indicates the SNR of each event. The vertical lines indicate where the SNR thresholds of 8 and 10 are located, respectively.
    }
    \label{fig:accumulated_logBF}
\end{figure}

As expected, these results indicate that the $(2,0)$ mode is too weak to yield statistically significant evidence for its presence in individual events at the current sensitivity.
We therefore perform a combined analysis by stacking the Bayes factors over all \GWTCFOUR BBH events.
We restrict this calculation to the \GWTCFOUR sample to avoid selection effects associated with including only the events selected for analysis from \GWTCFIVE. 
This approach enhances the overall sensitivity to a common signal component by coherently combining the individual evidences, allowing us to assess whether the data collectively support the presence of this mode beyond statistical fluctuations. 
The cumulative evidence trajectory is displayed in Fig.~\ref{fig:accumulated_logBF}. The events are sorted in growing network SNR order, with the corresponding value represented in the color bar. The shaded region shows the $1\sigma$ uncertainty accumulated over all the events.

The cumulative trajectory reaches a final value of $\log_{10} \mathcal{B}^{\mathrm{THM\_20/THM}} = 1.38 \pm 0.79$. Accounting for the uncertainty, the result does not provide significant evidence for the presence of the $(2,0)$ mode in the \GWTCFOUR sample. The high SNR events have the major contribution to the cumulative evidence. Although low SNR events contribute less to the cumulative Bayes factor than higher SNR events, the cumulative evidence continues to grow as additional events are included, indicating that the signal is not driven solely by a small number of high-significance detections.
Most detected events are compatible with face-on orientations of the binaries, where the dominant quadrupolar modes are strongest and the $(2,0)$ mode effectively vanishes. 
Overall, these results are consistent with previous studies under current detector sensitivity, 
as the $(2,0)$ mode represents a subdominant contribution relative to the primary signal, producing a weak per event imprint whose sign and magnitude depend on the SNR, and especially on parameters such as the inclination, the mass ratio and the spin components.

\begin{table}[htbp]
\centering
\begin{tabular}{cccc}
\hline\hline
Event & SNR & SNR$_{(2,0)}$ & $\log_{10}\mathcal{B}^{\mathrm{THM\_20/THM}}$ \\ 
\hline

GW230814\_230901 & $41.6^{+1.0}_{-1.0}$ & $0.44^{+0.39}_{-0.28}$ & $0.63\pm0.09$ \\
GW250114\_082203 & $75.5^{+1.0}_{-1.0}$ & $0.59^{+0.26}_{-0.23}$ & $0.30\pm0.12$ \\
GW230627\_015337 & $28.1^{+1.1}_{-1.1}$ & $0.49^{+0.30}_{-0.45}$ & $0.19\pm0.11$ \\
GW231224\_024321 & $12.4^{+1.1}_{-1.0}$ & $0.05^{+0.15}_{-0.04}$ & $0.13\pm0.10$ \\
GW240920\_124024 & $36.8^{+1.0}_{-1.0}$ & $0.10^{+0.15}_{-0.07}$ & $0.12\pm0.11$ \\
GW231129\_081745 & $7.1^{+1.1}_{-1.1}$ & $0.05^{+0.07}_{-0.04}$ & $0.10\pm0.08$ \\
GW230825\_041334 & $7.6^{+1.0}_{-1.1}$ & $0.02^{+0.05}_{-0.02}$ & $0.09\pm0.08$ \\
GW230704\_021211 & $8.6^{+1.0}_{-1.1}$ & $0.05^{+0.10}_{-0.04}$ & $0.09\pm0.08$ \\
GW230911\_195324 & $10.2^{+1.0}_{-1.1}$ & $0.10^{+0.07}_{-0.08}$ & $0.08\pm0.08$ \\
GW231123\_135430 & $19.3^{+1.0}_{-1.0}$ & $0.00^{+0.0}_{-0.01}$ & $-0.11\pm0.10$ \\
GW230914\_111401 & $15.8^{+1.0}_{-1.0}$ & $0.05^{+0.08}_{-0.04}$ & $-0.11\pm0.09$ \\
GW230708\_230935 & $8.7^{+1.0}_{-1.0}$ & $0.03^{+0.07}_{-0.03}$ & $-0.11\pm0.08$ \\
GW231014\_040532 & $8.2^{+1.0}_{-1.0}$ & $0.03^{+0.09}_{-0.03}$ & $-0.12\pm0.09$ \\
GW231226\_101520 & $33.3^{+1.1}_{-1.1}$ & $0.07^{+0.15}_{-0.06}$ & $-0.22\pm0.10$ \\

\hline\hline
\end{tabular}
\caption{List of events with log-10 Bayes factors above the selection threshold $\log_{10}\mathcal{B}^{\mathrm{THM\_20/THM}} - 1\sigma > 10^{-3}$ and below $\log_{10}\mathcal{B}^{\mathrm{THM\_20/THM}} + 1\sigma < -10^{-3}$. We include the SNR obtained with the \IMRPhenomTHMtwozero model for the full signal and for the $(2,0)$ mode. The quoted uncertainties represent one standard deviation. In Fig. \ref{fig:posteriors_events} we show the posteriors for some of the parameters of  the nine events with positive log-10 Bayes factors.}
\label{table:log10BF20_events}
\end{table}

\subsection{Stacking signal-to-noise ratio}
\label{sec:stackingSNR}

To compute the SNR of the $(2,0)$ mode for each event, we take the posterior samples of the parameters, we generate the corresponding waveform for each sample and compute the SNR of the mode, so that we obtain a posterior distribution of this quantity. In Fig. \ref{fig:posteriors_SNR20}, we show the posteriors for four events: GW230814\_230901, GW240925\_005809, GW241011\_233834 and GW250114\_082203. We do the same for the three options in the $(2,0)$ mode, so including only the displacement memory component, only the oscillatory component and both contributions. For most cases in the whole catalog, the median SNR of only the displacement memory is the lowest among the three contributions.
This is not the case for GW240925\_005809 and GW241011\_233834. GW240925\_005809 shows a bimodal distribution for the three posteriors. This bimodality is consistent with bimodality in the inferred redshift reported in \GWTCFIVE \cite{LIGOScientific:2026wfs,LIGOScientific:2026qqv}. As expected, the median for the full mode is the highest in all the events. For the rest of events not shown here, the SNR distributions rail against zero. However, for these four high SNR events, the posteriors deviate from zero, especially for GW250114\_082203, for which the median values of the three distributions are larger than zero, meaning that for this event the $(2,0)$ mode SNR is different from zero, but still below 1.

Once we have the estimation of this quantity for the three contributions and all events, we perform a stacking of the SNR using Eq. (\ref{eq:stackSNR}). The curves obtained for the three contributions and all the events in \GWTCFOUR are shown in Fig. \ref{fig:accumulated_SNR20} as a function of time (in years). For reference, we add a horizontal line at the threshold SNR=3, which is the threshold for memory used in \cite{Boersma:2020gxx,Grant:2022bla} and that we also adopt for comparison. This value is surpassed for the full $(2,0)$ mode, but not reached when we consider the individual components of this mode. The final values obtained from this calculation are summarized in Tab. \ref{table:SNR20finalvalues}, where we also include the final SNR reached when considering the selected events from \GWTCFIVE.

We now compare the light blue curve in Fig. \ref{fig:accumulated_SNR20} (SNR of the memory in the $(2,0)$) with the results presented in:
\begin{itemize}
    \item Fig. 4 from \cite{Boersma:2020gxx}: the curves show the total memory SNR with a detector network composed by Advanced LIGO and Advanced Virgo at design sensitivity.
    \item Fig. 5 from \cite{Grant:2022bla}: the blue curve corresponds to the displacement memory SNR estimate for the O4 network of detectors (LIGO Hanford, LIGO Livingston, Virgo, KAGRA and LIGO India) at design sensitivity.
\end{itemize}

The obtained results in this work are consistent with the prospects in these previous works, reaching a value which is near SNR=1 after $\sim7.5$ months of observations during O4a.

\begin{table}[htbp]
\centering
\begin{tabular}{cccl}
\hline\hline
Final value of (2,0) SNR stacking   & Full        & Oscillatory & Memory     \\
\hline
\GWTCFOUR events                 & $3.53^{+0.94}_{-0.41}$ & $2.57^{+0.66}_{-0.30}$ & $0.89^{+0.29}_{-0.11}$ \\
\GWTCFOUR + 6 \GWTCFIVE events & $4.93^{+1.04}_{-0.66}$ & $3.43^{+0.73}_{-0.39}$ & $1.40^{+0.34}_{-0.25}$ \\
\hline\hline
\end{tabular}
\caption{Final values of the cumulative SNR of the $(2,0)$ mode for each of the contributions in this mode and the full signal. Top row corresponds to the case where we only consider the events in \GWTCFOUR (see Fig. \ref{fig:accumulated_SNR20}), and the bottom row shows the results when adding the selected events from \GWTCFIVE.}
\label{table:SNR20finalvalues}
\end{table}

\begin{widetext}
\begin{center}
\begin{figure}[H]
    \centering
    \includegraphics[width=\textwidth]{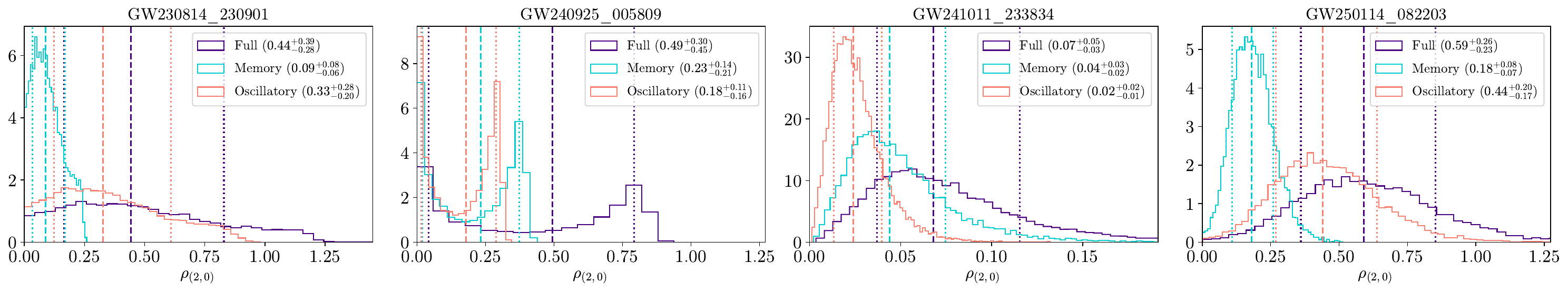}
    \caption{Posterior distributions of the SNR of the $(2,0)$ mode for four selected events. Each curve represents the distribution obtained with the different components in this mode. The numbers in the legend show the median, 16\% and 84\% percentiles of the distributions, denoted by the vertical lines in the plots (dashed-median, dotted-percentiles).
    }
    \label{fig:posteriors_SNR20}
\end{figure}
\end{center}
\end{widetext}

\begin{figure}[H]
    \centering
    \includegraphics[width=\columnwidth]{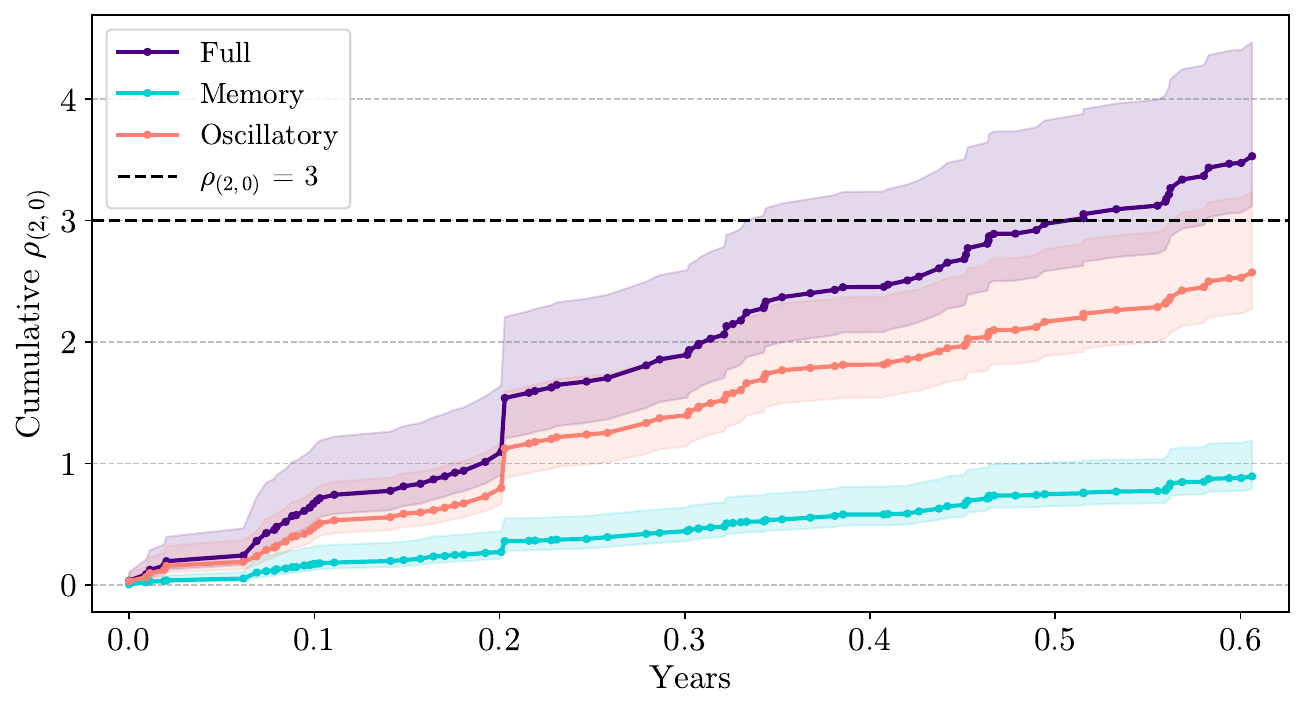}
    \caption{Cumulative sum of the SNR of the $(2,0)$ mode for the different contributions: the full mode, only the displacement memory and only the oscillatory components as a function of the time in years including only the events from \GWTCFOUR. The horizontal black, dashed line shows the threshold of SNR=3.
    }
    \label{fig:accumulated_SNR20}
\end{figure}

As expected, the SNR of the full $(2,0)$ mode accumulates more rapidly than either contribution considered separately, reaching a median value greater than 3 for the \GWTCFOUR event set. Since the \GWTCFIVE sample was selected to contain high SNR events, its inclusion in the final \GWTCFOUR results leads to a substantial increase in the corresponding values.

\subsection{GW241127\_061008: Analysis with spin-precessing models}
\label{sec:prec_event}
Since GW241127\_061008 shows clear evidence of spin precession \cite{LIGOScientific:2026wfs}, with a recovered value of $\chi_{\mathrm{p}} = 0.57^{+0.14}_{-0.21}$ that significantly deviates from zero, we analyze this event using the precessing version of the time domain waveform models. In particular, we compare \IMRPhenomTPHM, against the extended model, \IMRPhenomTPHMtwozero, which includes the $(2,0)$ mode in the co-precessing frame and the  displacement memory contribution in all the $\ell=2$ modes. This comparison allows us to assess whether these additional physical effects provide a measurable improvement in the description of the data. For the aligned-spin analysis, we obtain $\log_{10}\mathcal{B}^{\mathrm{THM\_20/THM}} = 0.05 \pm 0.12$, while for the precessing analysis we find $\log_{10}\mathcal{B}^{\mathrm{TPHM\_20/TPHM}} = -0.01 \pm 0.13$. These values indicate that, within the quoted uncertainties, the evidence in favor of including the additional $(2,0)$ mode and memory contributions is not statistically significant. 

\begin{figure}[H]
    \centering
    \includegraphics[width=\columnwidth]{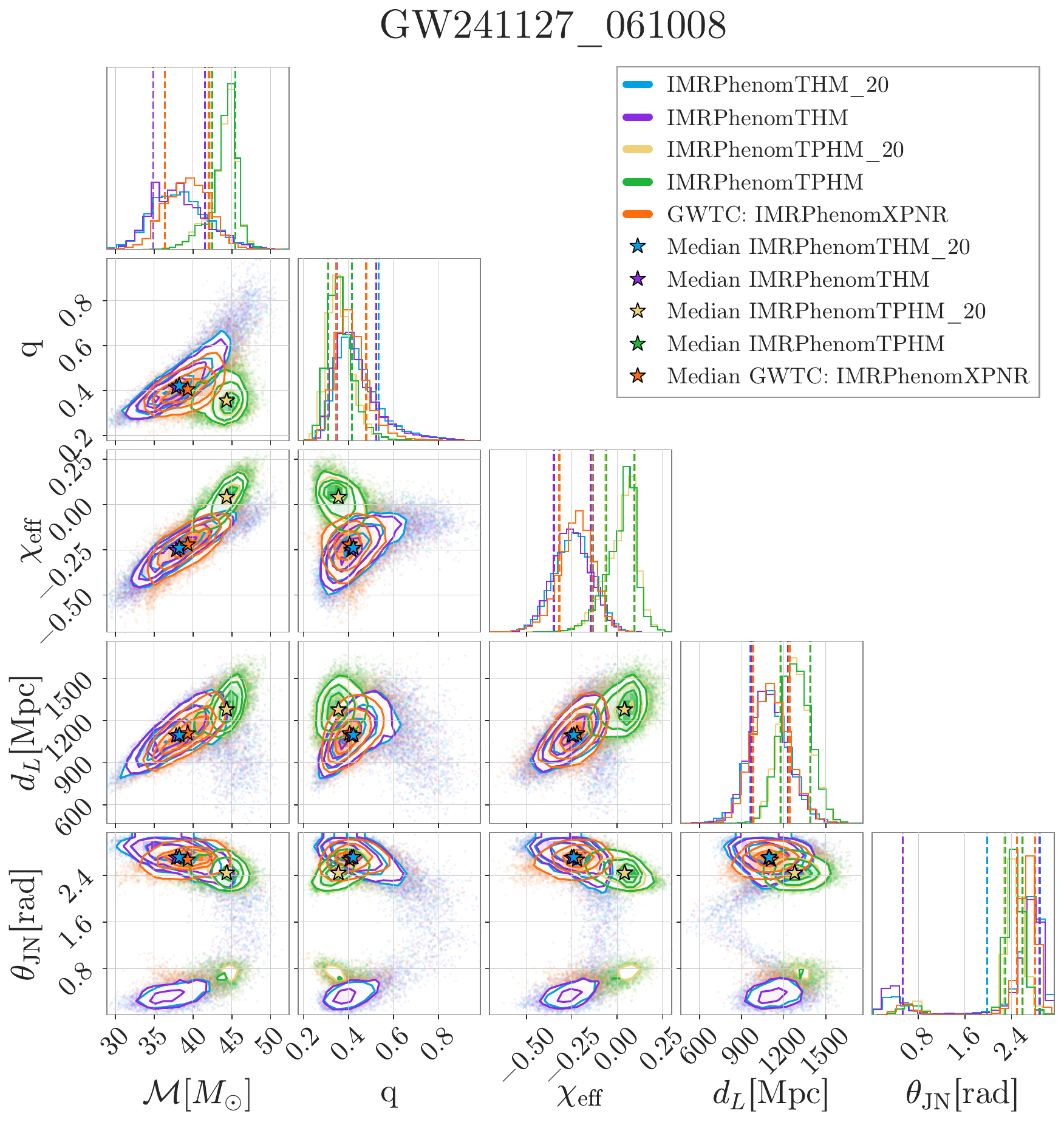}
    \caption{Posterior distributions for the event GW241127\_061008. The figure shows the parameters (chirp mass $\mathcal{M}$, mass ratio $q$, effective spin parameter $\chi_{\text{eff}}$, luminosity distance $d_L$ and inclination angle $\theta_{\text{JN}}$), recovered with \IMRPhenomTHMtwozero (blue), \IMRPhenomTHM (purple), \IMRPhenomTPHMtwozero (yellow), \IMRPhenomTPHM (green), and the samples from \GWTCFIVE obtained with \IMRPhenomXPNR (orange).}
    \label{fig:posterior_prec}
\end{figure}

In Fig.~\ref{fig:posterior_prec}, we show the posterior distributions for a set of parameters obtained with \IMRPhenomTHMtwozero, \IMRPhenomTHM, \IMRPhenomTPHMtwozero, and \IMRPhenomTPHM. For comparison, we also include the posterior samples from \IMRPhenomXPNR reported in \GWTCFIVE. Overall, we find that including the $(2,0)$ mode in both the aligned-spin and precessing versions of the model does not produce any noticeable shift in the recovered parameters. Since \IMRPhenomTPHM does not model mode asymmetries, isolating a subdominant mode is more challenging, as its imprint can be partially masked by these unmodeled asymmetries.
Some differences arise between \IMRPhenomTHM and \IMRPhenomTPHM: the precessing models recover smaller values of $q$ and a $\chi_{\text{eff}}$ distribution peaked near zero, whereas the aligned-spin models favor more unequal mass ratios and moderately negative values of $\chi_{\text{eff}}$, consistent with the fact that the $q$--$\chi_{\text{eff}}$ degeneracy is broken differently when precession is allowed. The precessing versions of the models also recover distributions shifted toward larger luminosity distances and chirp masses. Relative to \IMRPhenomXPNR, the results of the aligned-spin models are broadly consistent across all parameters, whereas \IMRPhenomTPHM and \IMRPhenomTPHMtwozero show larger discrepancies, yielding a larger chirp mass, a slightly more equal mass ratio, and a $\chi_{\mathrm{eff}}$ closer to zero than \IMRPhenomXPNR. This is not entirely unexpected, since \IMRPhenomXPNR includes mode asymmetries and additional precession effects beyond those modeled in \IMRPhenomTPHM. The inclination angle, $\theta_{\text{JN}}$, remains bimodal for all models.

\subsection{Detection prospects}
\label{sec:prospects}
Using state-of-the-art phenomenological waveform models calibrated to numerical relativity simulations this work has set constraints on the detectability of the $(2,0)$ mode in individual events, which remains a challenge in ground-based detectors. In the following we use the present analysis to make projections for the number of observations required to obtain clear evidence for this mode at the current detector sensitivity.

To model the empirical distributions of $\log_{10}\mathcal{B}^{\mathrm{THM\_20/THM}}$ and $\sigma[\log_{10}\mathcal{B}^{\mathrm{THM\_20/THM}}]$ inferred from the \GWTCFOUR BBH events, we fit parametric probability distributions to the observed samples. We adopt a Student’s $t$ distribution for $\log_{10}\mathcal{B}^{\mathrm{THM\_20/THM}}$, as the data exhibit heavier tails than a Gaussian, and a skew-normal distribution for the associated uncertainties, which display clear asymmetry. We further verify that these choices provide the best description of the data among the tested families, based on the Akaike and Bayesian information criteria (AIC and BIC). The distribution parameters (including the degrees of freedom and shape parameter) are obtained via maximum-likelihood estimation. In particular, the fitted Student’s $t$ distribution yields 2.85 degrees of freedom, reflecting the pronounced non-Gaussianity and heavy tailed nature of the Bayes factor distribution, while the skew-normal fit to the uncertainty distribution gives a shape parameter of 8.33, capturing its positive skewness.

\begin{figure}[H]
    \centering
    \includegraphics[width=\columnwidth]{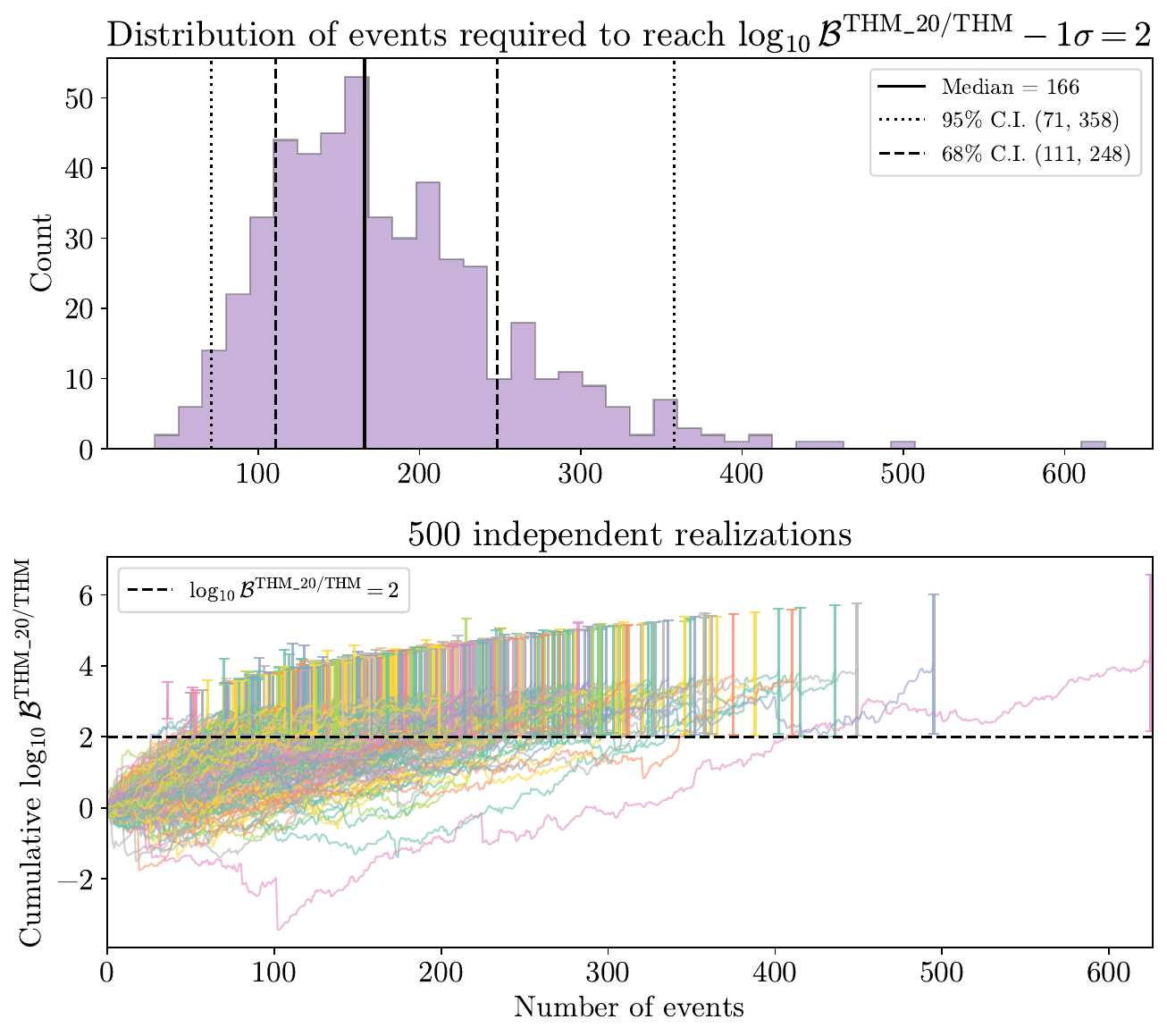}
    \caption{Distribution of the number of events required for the cumulative Bayes factor, $\log_{10}\mathcal{B}^{\mathrm{THM}\_20/\mathrm{THM}}$, to reach the threshold value of 2 accounting for the accumulated uncertainty, obtained from 500 independent realizations. The upper panel shows the histogram of threshold crossing event counts; the solid, dashed, and dotted vertical lines indicate the median, 68\% credible interval, and 95\% credible interval, respectively. The lower panel shows the cumulative evolution of $\log_{10}\mathcal{B}^{\mathrm{THM}\_20/\mathrm{THM}}$ as a function of the number of events for all realizations, with the horizontal dashed line marking the adopted threshold.}
    \label{fig:prospects}
\end{figure}

Random samples are then drawn from these distributions and accumulated until the cumulative sum reaches a fixed threshold, set to $\log_{10}\mathcal{B}^{\mathrm{THM\_20/THM}} - 1\sigma = 2$. This threshold is motivated by the Jeffreys scale, according to which values above 2 correspond to decisive statistical evidence. The procedure is repeated 500 times, yielding 500 independent realizations and, consequently, a distribution for the number of events required to reach the threshold. The corresponding results are shown in Fig.~\ref{fig:prospects}. We find that decisive evidence for the $(2,0)$ mode at the current detector sensitivity is expected to require a substantial catalog of detections, with an estimated number of events of $N_{\mathrm{events}} = 166^{+82}_{-55}$.
This estimate should be regarded as optimistic, as it is limited to the data considered in this work and to the assumptions adopted in the analysis.
A limitation of our work is that we do not account for waveform systematics, and the waveform model employed does not include the $(3,\pm2)$ mode. This omission may affect the Bayes factor estimates, as illustrated in particular by GW230814\_230901, for which the observed systematics appear to be driven primarily by the absence of this mode.
Moreover, the estimate is based solely on the set of \GWTCFOUR events, which may be too small a sample from which to infer robust distributional fits.
Nonetheless, under these assumptions, our results suggest that decisive evidence may be achievable by the end of O4, using the current BBH candidates reported from O4a and O4b (84 and 104 for a false alarm rate $>1\;\mathrm{yr}^{-1}$, respectively), and future candidates from O4c.


\section{conclusions}
\label{sec:conclusions}

In this work, we have carried out a systematic analysis 
of the BBH events in \GWTCFOUR, together with a selected subset of six high significance \GWTCFIVE events, using the \IMRPhenomTHM and \IMRPhenomTHMtwozero waveform models. The latter includes the full $(\ell=2,m=0)$ spherical harmonic mode, incorporating both the oscillatory ringdown contribution and the displacement memory component. Our analysis has combined event-by-event Bayesian model comparison with cumulative evidence accumulation and SNR stacking in order to assess the detectability of the $(2,0)$ mode in current ground-based GW data. 

At the level of individual events, the data do not provide statistically significant evidence for the inclusion of the $(2,0)$ mode in the majority of cases. The Bayes factors remain close to zero, with uncertainties that typically overlap the null value, indicating that the waveform models with and without this mode are generally indistinguishable at the current detector sensitivity. A mild trend toward larger Bayes factors is observed for louder events, and the strongest preference for the presence of the mode is obtained for GW230814\_230901. Importantly, the inclusion of the $(2,0)$ mode does not produce appreciable changes in the inferred source parameters for the events studied, which indicates that the omission of this mode is not yet a dominant source of systematic bias in current PE analyses. 

When the evidence is combined across the full O4a sample, we obtain a cumulative $\log_{10}\mathcal{B}^{\mathrm{THM\_20}/\mathrm{THM}} =1.38\pm0.79$, which, with its large uncertainty, constitutes only moderate support for the presence of the $(2,0)$ mode and falls short of decisive evidence according to the Jeffreys' scale. The stacking of the $(2,0)$ mode SNR further indicates that the full mode accumulates more rapidly than its separate oscillatory and memory contributions, reaching a median value above the commonly adopted threshold of $3$ for \GWTCFOUR, while the memory contribution alone remains below this level. For the precessing event GW241127\_061008, the comparison between the precessing waveform models similarly yields no statistically significant preference for including the additional $(2,0)$ contribution. 

These results suggest that the $(2,0)$ mode is beginning to contribute measurably to the loudest events in the current catalog, but that robust event level detection remains out of reach for present ground-based sensitivities. Our prospects analysis indicates that decisive evidence for the mode may only require a moderately larger catalog, with an estimated number of events of $N_{\mathrm{events}} = 166^{+82}_{-55}$. The main shortcoming of this first cumulative analysis with the full $(2,0)$ mode is that we have not studied effects of waveform systematics. While an analysis such as the one presented here can reveal very small effects such as the contributions of the $(2,0)$ mode, it can also be sensitive to systematics. Sufficient support for a robust claim of detection will require such systematics studies and more extended injection studies beyond the zero noise simplification.

Beyond the observational implications for the $(2,0)$ mode, this work also constitutes a step toward the detection of GW displacement memory. Given the additional SNR that is obtained from the oscillatory component of the $(2,0)$ mode, we foresee that decisive evidence for the entire mode will be obtained before a clear detection of the separate memory component.
We expect that controlling the presence of the full $(2,0)$ mode will support memory detection also in terms of controlling waveform systematics, in particular to exclude contamination from systematics effects in the part of waveform that is not the memory.

\section*{Acknowledgements}

The authors would like to thank Shubhanshu Tiwari for the LSC Publication \& Presentation Committee review of this manuscript. We thank Héctor Estellés for useful discussions and suggestions. This paper has document number LIGO-P2600341.
We thankfully acknowledge the computer resources (MN5 Supercomputer), technical expertise and assistance provided by Barcelona Supercomputing Center (BSC)  through funding from the Red Española de Supercomputación (RES) (AECT-2024-3-0027); and the computer resources (Picasso Supercomputer), technical expertise and assistance provided by the SCBI (Supercomputing and Bioinformatics) center of the University of Málaga (AECT-2025-1-0035).
This research has made use of data or software obtained from the Gravitational Wave Open Science Center (gwosc.org), a service of the LIGO Scientific Collaboration, the Virgo Collaboration, and KAGRA.
This material is based upon work supported by NSF's LIGO Laboratory which is a major facility fully funded by the National Science Foundation.
LIGO is funded by the U.S. National Science Foundation. Virgo is funded by the French Centre National de Recherche Scientifique (CNRS), the Italian Istituto Nazionale della Fisica Nucleare (INFN) and the Dutch Nikhef, with contributions by Polish and Hungarian institutes.
M. Rosselló-Sastre and J. Valencia were supported by the Spanish Ministry of Universities, grants FPU21/05009 and FPU22/02211, respectively.
Y. Xu was supported by the INVESTIGA@UIB programme of the Universitat de les Illes Balears (UIB), co-funded by the 2023 Sustainable Tourism Promotion Plan (ITS2023-086 – Research Promotion Programme).
J. Llobera-Querol was supported by the Conselleria d'Educació i Universitats del Govern de les Illes Balears via FPI-CAIB doctoral grant FPI\_093\_2022.
A. Ramos-Buades is supported by the Veni research programme which is (partly) financed by the Dutch Research Council (NWO) under the grant VI.Veni.222.396; 
acknowledges support from the Spanish Agencia Estatal de Investigación grant PID2024-157460NA-I00; and the Spanish Ministerio de Ciencia, Innovación y Universidades (Beatriz Galindo, BG23/00056), co-financed by UIB.

This work was supported by the Universitat de les Illes Balears (UIB) with funds from the Programa de Foment de la Recerca i la Innovació de la UIB 2024-2026 (supported by the yearly plan of the Tourist Stay Tax ITS2023-086); the Spanish Agencia Estatal de Investigación grants PID2022-138626NB-I00, RED2024-153978-E, RED2024-153735-E, funded by MICIU/AEI/10.13039/501100011033 and the ERDF/EU; and the Comunitat Autònoma de les Illes Balears through the Conselleria d'Educació i Universitats with funds from the European Union - European Regional Development Fund (ERDF) (SINCO2022/18146 - Plataforma HiTech-IAC3-BIO). 

\clearpage

\begin{widetext}

\appendix

\section{Computational cost of the runs when varying $\Delta\log Z$}
\label{app:runtimes_dlogz}
In Fig.~\ref{fig:runtimes}, we show the relative runtime difference, in percent, between analyses performed with the default setting, $\Delta\log Z=0.10$, and with half that value, $\Delta\log Z=0.05$, for each GW event and waveform model version. Markers above (below) 0 indicate that the runtime for the reduced threshold is higher (lower) than for the default setting. The color bar shows the absolute runtime, in hours, obtained for the default setting, providing a reference for the computational cost associated with each event. The largest relative differences are found for short duration events, indicating that lowering $\Delta\log Z$ does not necessarily lead to a substantial increase in the computational cost of the analysis. All runs were performed on the Picasso supercomputer using 128 CPU cores. The reported runtimes therefore correspond to this computing configuration and may vary for different hardware resources.

\begin{figure}[H]
    \centering
    \includegraphics[width=\columnwidth]{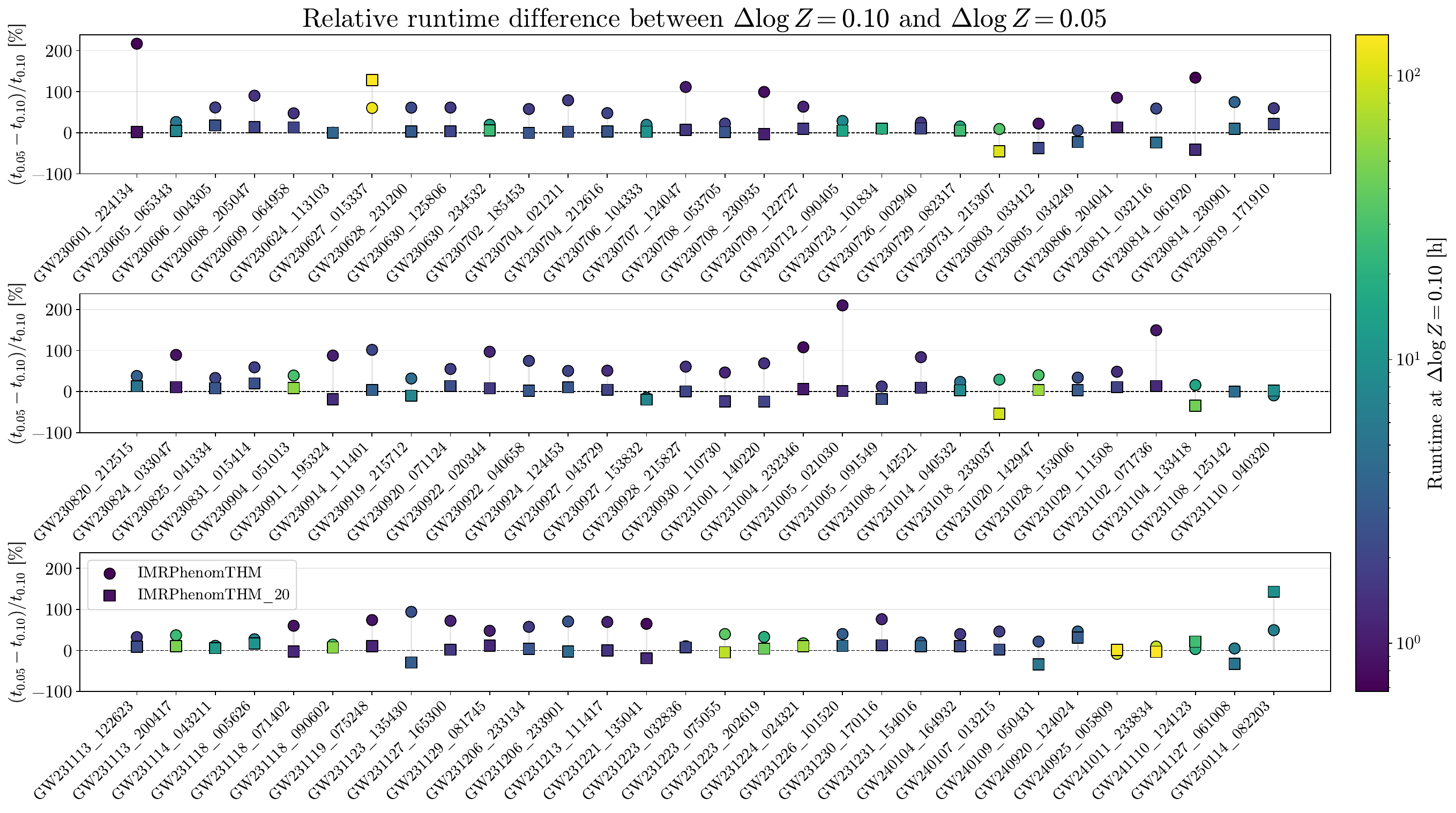}
    \caption{Relative runtime difference between the runs performed with $\Delta\log Z=0.10$ and $\Delta\log Z=0.05$ for all analyzed events. The vertical axis shows the relative difference in runtime, defined as $(t_{0.05}-t_{0.10})/t_{0.10}$ in percent, where positive values indicate longer runtimes for the more stringent stopping criterion. Circles and squares correspond to the \IMRPhenomTHM and \IMRPhenomTHMtwozero models, respectively. The marker color indicates the absolute runtime obtained with $\Delta\log Z=0.10$ (in hours), as shown by the color bar on the right. The events are chronologically ordered.}
    \label{fig:runtimes}
\end{figure}

\clearpage
\section{Numerical Relativity injection}
\label{app:nrinjection}
We present in Fig.~\ref{fig:posterior_NRinjection} the zero-noise NR injection with the simulation SXS:BBH:2497 ($Q=2,\chi_{1}=\chi_{2}=0, N_{\mathrm{orbits}}=20$) recovering with \IMRPhenomTHM and \IMRPhenomTHMtwozero. 
As discussed in Sec.~\ref{sec:injections}, this injection study is performed as a consistency check to validate the behavior observed in the analysis of real data. We find that the parameter biases introduced by neglecting the $(2,0)$ mode are generally negligible, with only marginal effects observed across most of the recovered source parameters.

\begin{figure}[H]
    \centering
    \includegraphics[width=0.98\textwidth]{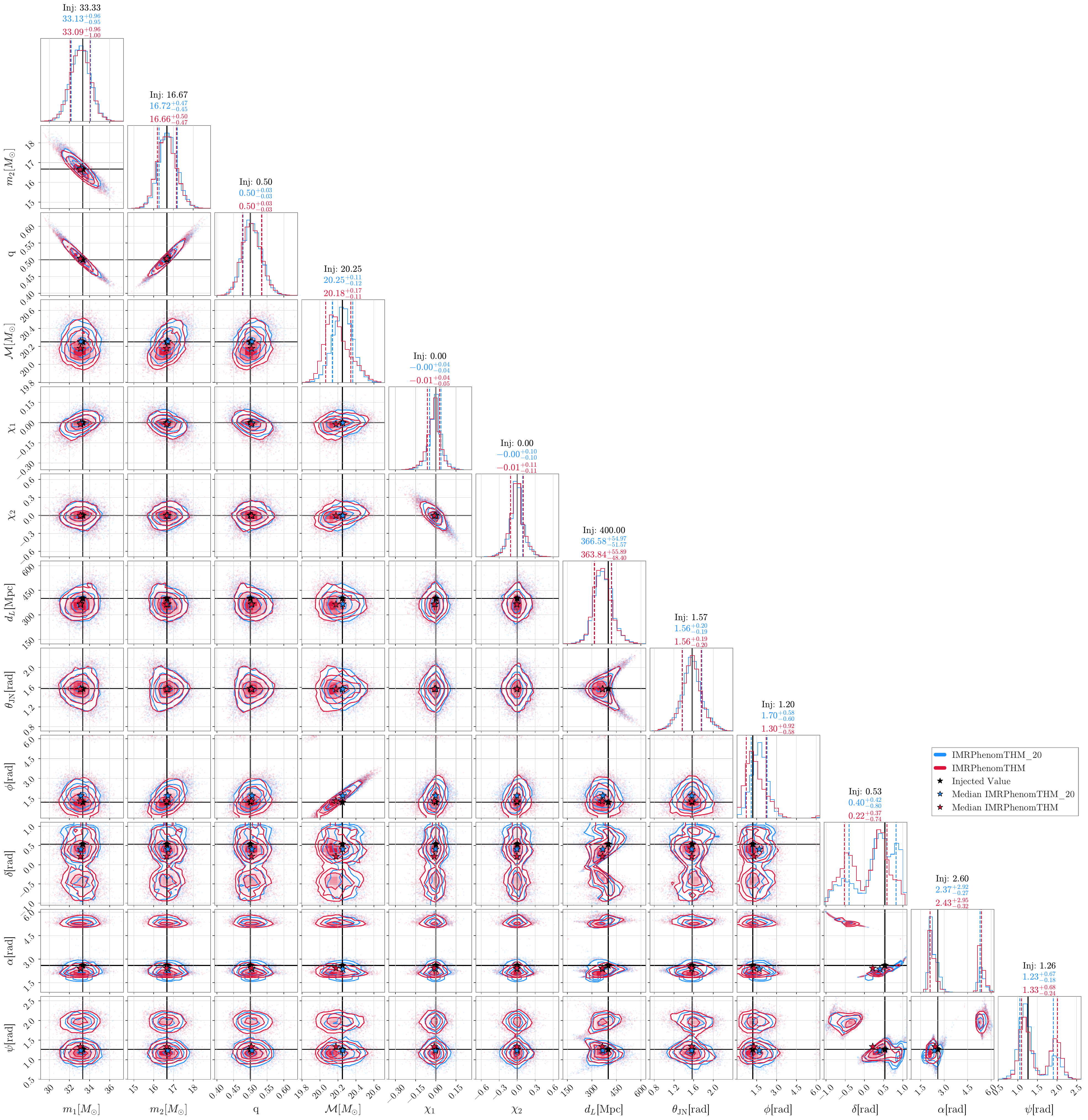}
    \caption{Corner plot 
    for the NR injection with SXS:BBH:2497, recovering with \IMRPhenomTHMtwozero (blue) and \IMRPhenomTHM (red), showing 
    the 2D and 1D posteriors of the individual masses $m_1$ and $m_2$, mass ratio $q$, chirp mass $\mathcal{M}$, individual spin components aligned with the orbital angular momentum $\chi_{1}$ and $\chi_{2}$, luminosity distance $d_L$, inclination $\theta_{\text{JN}}$, phase $\phi$, declination $\delta$, right ascension $\alpha$ and polarization angle $\psi$. The injected values are shown in black. 
    }
    \label{fig:posterior_NRinjection}
\end{figure}

\newpage
\section{Further results for the events with highest $\log_{10}\mathcal{B}^{\text{THM\_20/THM}}$}
\label{app:posteriors}
Fig.~\ref{fig:posteriors_events} shows the posterior distributions of selected source parameters for the nine events exhibiting the highest statistical evidence for the presence of the $(2,0)$ mode. For each event, we compare the parameter recovery obtained with waveform models including and excluding this mode in order to assess potential systematic biases arising from its omission. In all cases, the recovered posterior distributions are found to be largely consistent between the two models, indicating that neglecting the $(2,0)$ mode produces only negligible effects on the inferred source parameters. For reference, we also include the posteriors reported by the LVK in the corresponding GWTCs.

\begin{figure}[H]
    \centering
    \includegraphics[width=0.9\textwidth]{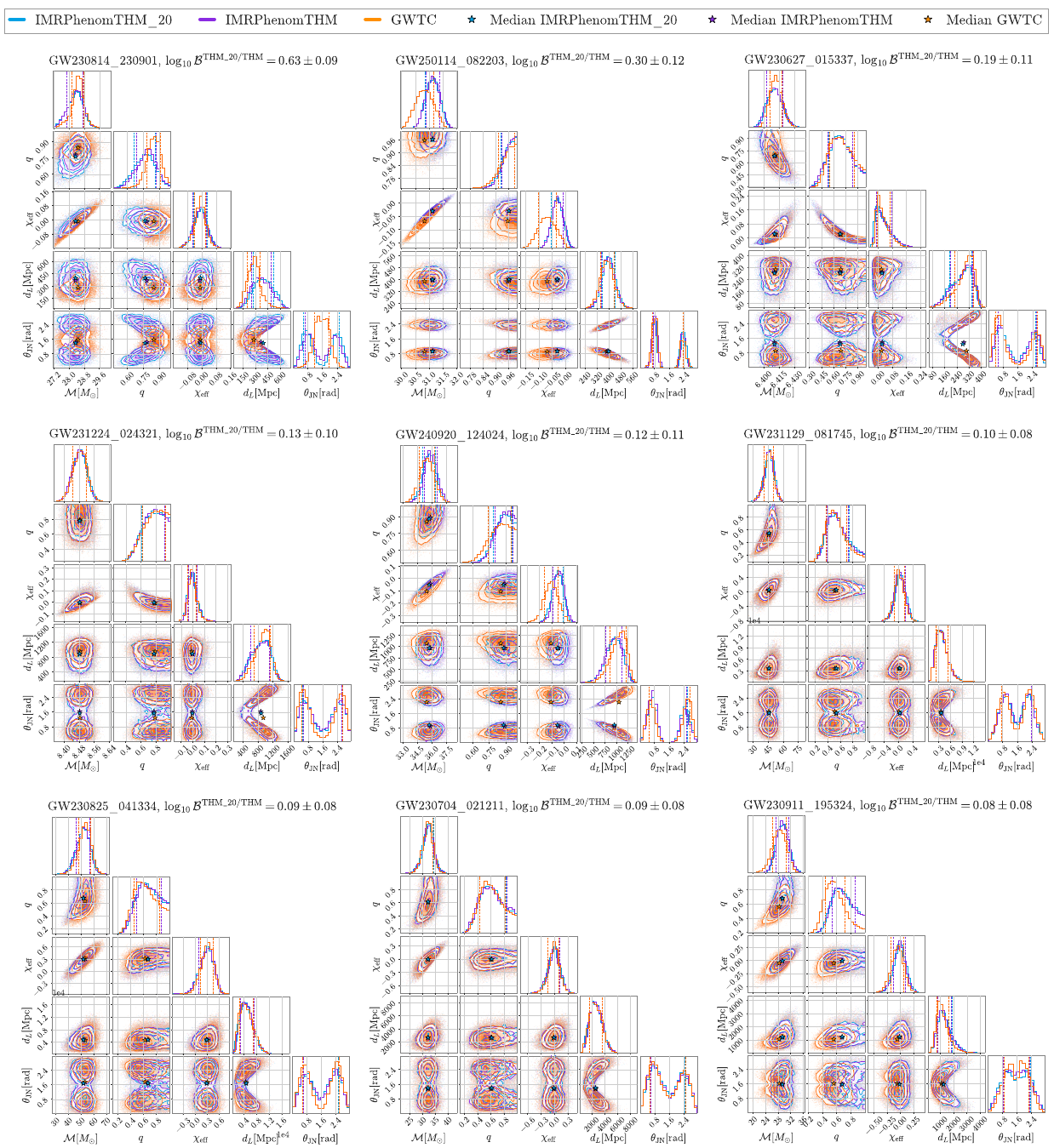}
    \caption{Posterior distributions for the events listed in Tab. \ref{table:log10BF20_events} satisfying $\log_{10}\mathcal{B}^{\mathrm{THM\_20/THM}} - 1\sigma > 10^{-3}$. The figure shows the parameters (chirp mass $\mathcal{M}$, mass ratio $q$, effective spin parameter $\chi_{\text{eff}}$, luminosity distance $d_L$ and inclination $\theta_{\text{JN}}$), recovered with \IMRPhenomTHMtwozero (blue), \IMRPhenomTHM (purple), and the samples from GWTC (orange).
    }
    \label{fig:posteriors_events}
\end{figure}

In Fig.~\ref{fig:whitened_WFs}, we show the whitened waveforms of GW230814\_230901 (single detector event, L1 data shown in the top panel) and GW250114\_082203 (H1 data in the middle panel and L1 data in the bottom panel). In these events, the signal morphology is clearly distinguishable from the detector noise, enabling a direct visual comparison between models and data. In both cases, the maximum-likelihood waveforms obtained from the two models provide a good fit to the data. The differences between the two waveform models with and without the $(2,0)$ mode 
are small and are shown by the dashed black curves. These discrepancies become more pronounced in the late inspiral and merger phase, consistent with the expectation that the signal contribution of the $(2,0)$ mode is predominantly concentrated in this regime. The absence of any apparent memory induced offset in the whitened waveforms arises as a direct consequence of the whitening procedure. By dividing the strain by the noise PSD, whitening effectively acts as a high pass filter that suppresses the low frequency content of the signal. As a result, the low frequency components associated with the memory are strongly attenuated, removing any constant strain offset and leaving only the oscillatory structure of the waveform.

\begin{figure}[H]
    \centering
    \includegraphics[width=0.76\textwidth]{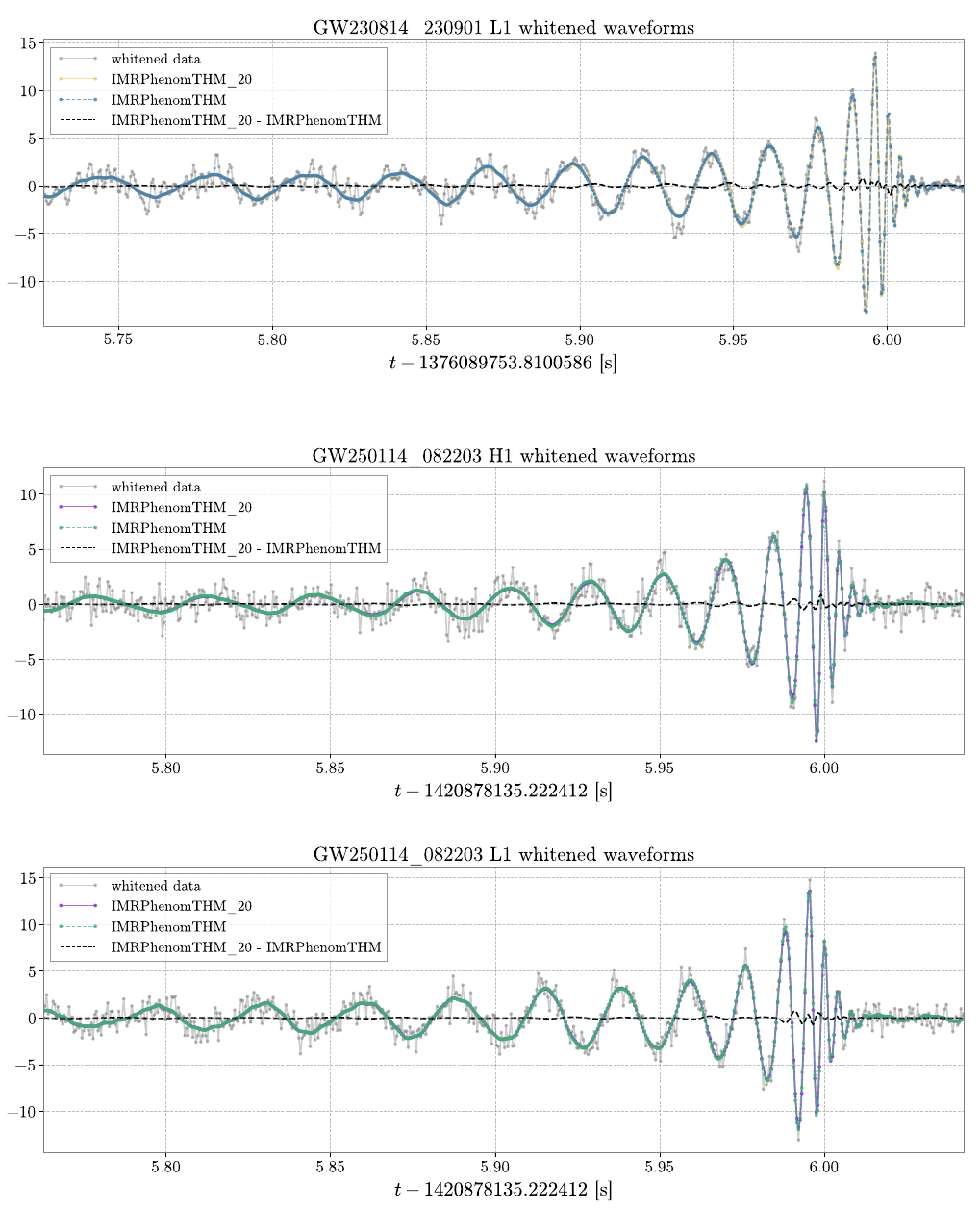}
    \caption{Whitened waveforms for GW230814\_230901 (only detected in L1) and GW250114\_082203 (detected in both L1 and H1). The gray lines show the whitened data, while the colored lines show the maximum likelihood waveforms for each version of the model. In black we include the difference between the two waveform models.}
    \label{fig:whitened_WFs}
\end{figure}

\end{widetext}




\let\c\Originalcdefinition %
\let\d\Originalddefinition %
\let\i\Originalidefinition

\bibliography{bib}

\end{document}

%% file: UsePackages.tex
\usepackage[T3,T1]{fontenc}
\usepackage{mathrsfs} 
\usepackage{bm} 
\makeatletter
\@ifclassloaded{beamer}
  { 
    \typeout{UsePackages: Detected beamer}
    \usepackage{tgheros}
    
  }
  { 
    \typeout{UsePackages: Did not detect beamer}
    \ifx\asybeamer\undefined 
    \typeout{UsePackages: Detected article}
    \usepackage[varg]{txfonts} 
    \usepackage{tgtermes} 
    \else 
    \typeout{Fonts: Detected asy for beamer}
    \usepackage{tgheros}
    
    \fi
  }
\usepackage{microtype} 
\def\MT@register@subst@font{
  \MT@exp@one@n\MT@in@clist\font@name\MT@font@list
  \ifMT@inlist@\else\xdef\MT@font@list{\MT@font@list\font@name,}\fi}
\makeatother

\usepackage{graphicx} %
\usepackage[x11names,svgnames,rgb]{xcolor} %
\usepackage{xspace}
\usepackage{braket}
\usepackage{accents}
\usepackage{siunitx}
\usepackage{mathtools}
\DeclareSymbolFontAlphabet{\mathrm}{operators}

\definecolor{CiteColor}{rgb}{0.18039, 0.18824, 0.57255}
\definecolor{UrlColor} {rgb}{0.741, 0.173, 0.000}
\definecolor{DarkUrlColor} {rgb}{0.500, 0.110, 0.000}
\definecolor{LinkColor}{rgb}{0.25098, 0.47843, 0.04706}

\makeatletter %
\newcommand{\ShowFont}{%
  \typeout{The main font is \f@encoding \space \f@family \space %
    \f@series \space \f@shape \space at \f@size pt.}%
  \typeout{The math font sizes are \tf@size pt (main), \sf@size pt %
    (script), and \ssf@size pt (scriptscript).}%
  \typeout{The linewidth is \the\linewidth}} %
\makeatother %

\usepackage{xargs}
